\documentclass[aps, pra, reprint, floatfix]{revtex4-2}

\usepackage[dvips]{graphicx}
\usepackage{epsfig}

\usepackage[margin=2cm,head=0.5cm]{geometry}
\usepackage{bm}
\usepackage{amsmath}
\usepackage{amssymb}
\usepackage{latexsym}
\usepackage{amsfonts}
\usepackage{epsfig}
\usepackage{color}
\usepackage[linktocpage, colorlinks=true ,linkcolor=blue, citecolor=blue]{hyperref}
\usepackage[all]{hypcap}

\newcommand{\expval}[1]{\left< #1 \right>}

\newcommand{\nn}{\nonumber\\}

\newcommand{\f}[1]{\mbox{\boldmath$#1$}}

\newcommand{\bea}{\begin{eqnarray}}
\newcommand{\eea}{\end{eqnarray}}
\newcommand{\beann}{\begin{eqnarray*}}
\newcommand{\eeann}{\end{eqnarray*}}
\newcommand{\ord}{{\cal O}}
\newcommand{\trace}[1]{{\rm Tr}\left\{ #1 \right\}}

\newcommand{\abs}[1]{{\left| #1 \right|}}

\newcommand{\ii}{\mathrm{i}}  

\begin{document}
  
\title{Exploring bosonic bound states with parallel reaction coordinates}
\author{Guan-Yu Lai}
\author{Friedemann Quei{\ss}er}
\affiliation{Helmholtz-Zentrum Dresden-Rossendorf, Bautzner Landstraße 400, 01328 Dresden, Germany}
\author{Gernot Schaller}
\email{g.schaller@hzdr.de}
\affiliation{Helmholtz-Zentrum Dresden-Rossendorf, Bautzner Landstraße 400, 01328 Dresden, Germany}
\date{\today}
\begin{abstract}
Bound states are dissipation-resilient states that may emerge when quantum systems are strongly coupled to reservoirs with band gaps. 
We analyze an exactly solvable bosonic model for bound state existence and reproduce these
results by a weak-coupling treatment of a supersystem composed of the original system and multiple reaction coordinates,
which are individually representing small energy intervals of the reservoir spectral function.
Within the perturbative supersystem treatment, the bound state stability results from its energy being inside the band gap.
We discuss cases of multiple band gaps and also show that already in presence of weak interactions the bound state's lifetime is finite -- but can be increased by 
raising the system-reservoir coupling strength.
\end{abstract}

\maketitle

\section{Introduction}

When you bring a quantum system in contact with a reservoir, its fragile quantum information will typically quickly and irreversibly disperse over the reservoir degrees of freedom~\cite{breuer2007}.
The specifics of this information loss can be used to classify the relaxation process as Markovian or non-Markovian~\cite{wolf2008a,breuer2009c,rivas2010a,breuer2012a}, respectively, but in the long run, it completely decoheres the system.
This process is responsible for our inability to control quantum systems well and is a significant obstacle~\cite{unruh1995a,reina2002a} to the construction of a scalable quantum computer~\cite{divincenzo2000a,nielsen2000}.

Therefore, it is highly intriguing that some quantum states may be robust to the influence a continuous reservoir -- even in absence of symmetries and at large ambient temperature.
When they arise from the (single-particle) band structure of the reservoirs, they are called bound states (BSs) or localized modes~\cite{mahan1990,marinica2008a,hsu2016a}.
Their existence is independent of the reservoir statistics~\cite{longhi2007a}, and consequently they have been studied both in bosonic~\cite{john1990a,kofman1994a,angelakis2004a,chang2018a} 
and fermionic~\cite{dhar2006a,stefanucci2007a,jussiau2019a} reservoirs.
Experimental observations have also been reported~\cite{plotnik2011a,amrani2021a}.
They can emerge when the spectral function of the reservoir exhibits band gaps, i.e., when it has regions where it strictly vanishes.
However, this alone is not a sufficient condition for BS existence.
Additionally, the system-reservoir coupling strength has to be strong enough, a criterion that usually forbids the use of perturbative schemes.
Typically, BSs are discussed for integrable systems.

In this paper, we generalize the reaction-coordinate (RC) mapping~\cite{martinazzo2011a,woods2014a,strasberg2016a,nazir2019a,correa2019a,anto_sztrikacs2021a} to explore the asymptotic long-term dynamics of the BS.
We start in Sec.~\ref{SEC:model} by introducing our example model and review the basic characteristics of its exact long-term solution in Sec.~\ref{SEC:exactsolution}.
We then introduce the details of the RC mapping and compare with the previous results in Sec.~\ref{SEC:rcmapping}.
After discussing the effects of multiple bands and interactions in Sec.~\ref{SEC:extensions} we conclude in Sec.~\ref{SEC:summary}.
Technical details are provided in several appendices.

\section{Model}\label{SEC:model}

\begin{figure}[ht]
\includegraphics[width=0.5\textwidth]{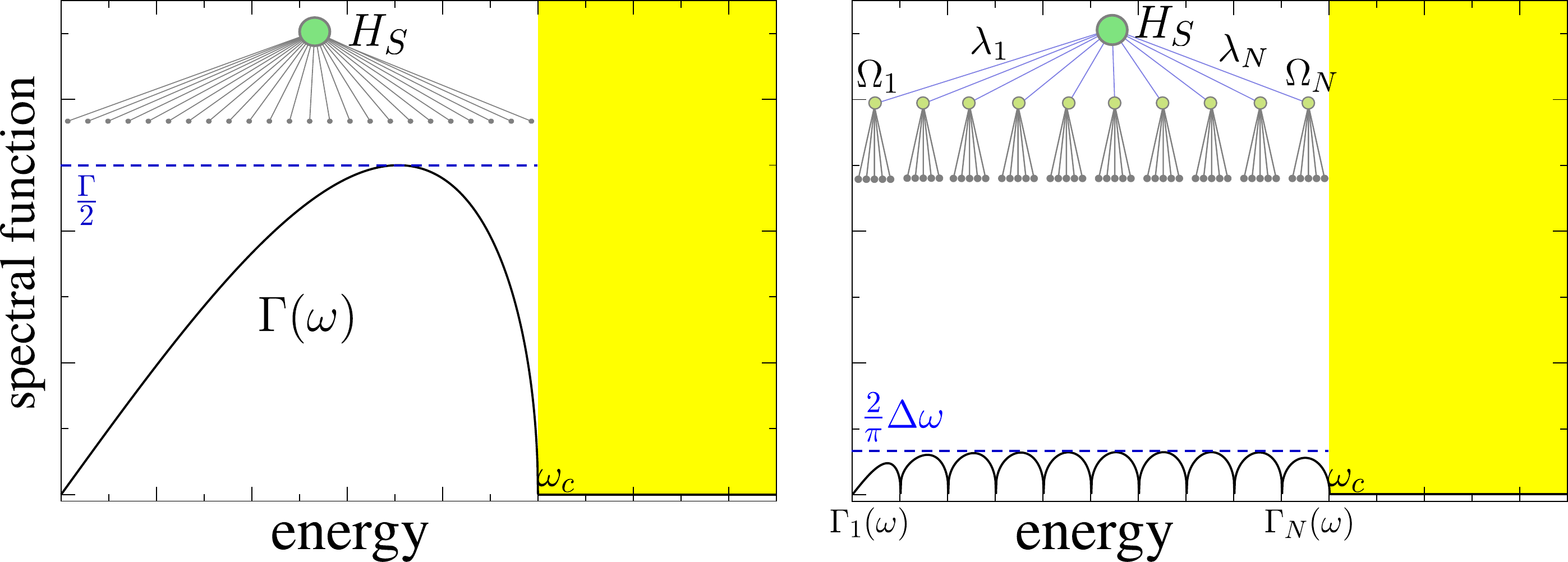}
\caption{\label{FIG:rcmappingspecdens}
{\bf Left:} The system is coupled to the reservoir-modes in a star-shaped fashion (inset), with the coupling strength characterized by the spectral function $\Gamma(\omega)$.
{\bf Right:} Partitioning the reservoir energy range into (not necessarily equal-size) disjoint intervals, we perform a reaction-coordinate (RC) mapping for each interval, leading to couplings $\lambda_i$ and RC energies $\Omega_i$. For fine discretizations, the residual spectral functions $\Gamma_i(\omega)$ for the $i$th RC approach an asymptotic bound (dashed blue) that only depends on discretization interval size, but the overall band gap (yellow) remains the same.
}
\end{figure}

Our model consists of a bosonic mode $a$ in the system that is coupled to bosonic reservoir modes $b_k$ via the amplitudes $h_k$ (compare Fig.~\ref{FIG:rcmappingspecdens} left panel)
\begin{align}\label{EQ:model1}
H = H_S + \sum_k \omega_k \left[b_k^\dagger + \frac{h_k}{\omega_k}(a+a^\dagger)\right]\left[b_k + \frac{h_k^*}{\omega_k}(a+a^\dagger)\right]\,.
\end{align}
The total Hamiltonian has a lower spectral bound for all values of $h_k$ when the system Hamiltonian $H_S$ is lower-bounded (which we assume throughout).
In the continuum limit, the eigenvalues $\omega_k$ of reservoir modes become dense and we can introduce the spectral function (spectral coupling density)~\cite{leggett1987a}
\begin{align}
\Gamma(\omega) = 2\pi \sum_k \abs{h_k}^2 \delta(\omega-\omega_k)
\end{align}
that quantifies the energy dependence of the system-reservoir coupling strength.
Loosely speaking, a perturbative scheme can be applied when $\Gamma(\omega)$ is small.
In Eq.~\eqref{EQ:model1}, the Hamiltonian assumes a star-shaped configuration (which can for quadratic models always be achieved by diagonalizing the reservoir), since the reservoir 
modes only interact with each other indirectly via the system, see also the inset in Fig.~\ref{FIG:rcmappingspecdens} left panel.

\section{Exact solution}\label{SEC:exactsolution}

Taking the system oscillator as harmonic
\begin{align}
H_S=\Omega a^\dagger a\,,
\end{align}
the model~\eqref{EQ:model1} remains quadratic, which implies that it is exactly solvable.
One approach relies on solving the Heisenberg equations of motion for the operators $\f{a}=e^{+\ii H t} a e^{-\ii H t}$ and $\f{b_k} = e^{+\ii H t} b_k e^{-\ii H t}$ and their hermitian conjugates with a Laplace transform~\cite{schaller2009a,zhang2012a,topp2015a,jussiau2019a}.
By identifying the purely imaginary poles of the solution in the Laplace domain it is possible to obtain the asymptotic long-term dynamics, see App.~\ref{APP:exactsolution}.
There, one observes fundamentally different behaviour for spectral functions that have no band gaps as e.g. $\Gamma(\omega) = \Gamma \frac{\omega\omega_c}{\omega^2+\omega_c^2}\Theta(\omega)$ vs. gapped ones like the Rubin spectral function (describing the coupling to a semi-infinite oscillator chain~\cite{rubin1963a,weiss1993})
\begin{align}\label{EQ:sfrubin}
\Gamma(\omega) = \Gamma \frac{\omega}{\omega_c} \sqrt{1-\frac{\omega^2}{\omega_c^2}}\Theta(\omega)\Theta(\omega_c-\omega)\,,
\end{align}
which strictly vanishes for $\omega>\omega_c$, compare Fig.~\ref{FIG:rcmappingspecdens} left panel.
For this spectral function, we find that in the long-term limit and strong system-reservoir couplings the position of the oscillator and its second moment may conditionally maintain an oscillatory motion {\em ad infinitum} -- a BS signature.
Particularly, when 
\begin{align}\label{EQ:boundstatecondition}
1 - \frac{\Omega^2}{\omega_c^2} \le \frac{\Gamma\Omega}{\omega_c^2}\,,
\end{align}
we determine the frequency of the BS as
\begin{align}\label{EQ:prediction_boundstate}
\omega_b = \omega_c \sqrt{\frac{\alpha^2+2\alpha\beta^2-2\beta^2+\alpha\sqrt{(\alpha+2\beta^2)^2-4\beta^2}}{4\alpha-2}}
\end{align}
with $\alpha=\Gamma\Omega/\omega_c^2$ and $\beta=\Omega/\omega_c$, such that one has $\omega_b>\omega_c$.
Then, the observables evolve in the long-term limit asymptotically as
\begin{align}\label{EQ:xsolution_time}
\expval{\f{x}}_\infty &= g(t) \expval{x}_0 + \frac{\Omega}{\omega_b} h(t) \expval{p}_0\,,\nn
\expval{\f{x}^2}_\infty &= \expval{\left[g(t) x + \frac{\Omega}{\omega_b} h(t) p\right]^2}_0\nn
&\qquad+\int\limits_0^{\omega_c}\frac{d\omega}{2\pi} 
\frac{\Gamma(\omega)[1+2n(\omega)] 4\Omega^2 [g^2(t)+\frac{\omega^2}{\omega_b^2} h^2(t)]}{(\omega_b^2-\omega^2)^2}\nn
&\qquad+\int\limits_{0}^{\omega_c}\frac{d\omega}{2\pi} \frac{\Gamma(\omega)[1+2 n(\omega)]4\Omega^2}{[\omega^2-\Omega f(-\ii\omega)][\omega^2-\Omega f(\ii\omega)]}\,,\nn
g(t) &= \frac{\cos(\omega_b t)}{1+\frac{\Omega\bar f}{2\omega_b}}\,,\qquad
h(t) = \frac{\sin(\omega_b t)}{1+\frac{\Omega\bar f}{2\omega_b}}\,,
\end{align}
with Bose distribution $n(\omega)=[e^{\omega/(k_B T)}-1]^{-1}$ and 
\begin{align}
\bar f &= \frac{4\omega_b}{\pi} \int\limits_0^{\omega_c} \frac{\Gamma(\omega)\omega d\omega }{(\omega^2-\omega_b^2)^2}
\stackrel{\small{\omega_b > \omega_c}}{=} \frac{\Gamma \omega_b \left[1-\sqrt{1-\frac{\omega_c^2}{\omega_b^2}}\right]^2}{\omega_c^2 \sqrt{1-\frac{\omega_c^2}{\omega_b^2}}}\,,\nn
f(\pm\ii\omega) &= \Omega + \frac{2}{\pi} I_{\cal P}(\omega) \pm\ii\Gamma(\omega)\,,\nn
I_{\cal P}(\omega) &= {\cal P}\int_0^{\omega_c} \frac{\Gamma(\bar\omega)}{\bar\omega} \frac{\omega^2}{\omega^2-\bar\omega^2} d\bar\omega = \frac{\Gamma \pi}{2} \frac{\omega^2}{\omega_c^2}\,.
\end{align}
Note that $\bar f$ is finite when $\omega_b > \omega_c$.
In contrast, in case condition~\eqref{EQ:boundstatecondition} is not fulfilled, we have to formally take $g(t),h(t)\to 0$, i.e., in Eq.~\eqref{EQ:xsolution_time} the first moment vanishes 
and for the second moment only the last continuum contribution remains.
When we additionally consider the weak-coupling regime, where $\Gamma(\omega)\to 0$, we obtain a Dirac-$\delta$ function in the integrand, and the system thermalizes with the reservoir temperature $\expval{\f{x}^2}_\infty \to 1+2n(\Omega)$.
Analogous results hold for the momentum (see App.~\ref{APP:exactsolution}), and from $a^\dagger a = (x^2+p^2)/4-1/2$ we can also conclude the long-term occupation.
In absence of a BS (denoted by an overbar), the occupation becomes
\begin{align}
\expval{\overline{\f{a^\dagger a}}}_\infty &= \int_0^{\omega_c} \frac{d\omega}{2\pi} \frac{\Gamma(\omega)[1+2n(\omega)](\Omega^2+\omega^2)}{\abs{\omega^2-\Omega f(-\ii\omega)}^2} - \frac{1}{2}\,.
\end{align}

To summarize, provided $\Omega\in(0,\omega_c)$, our system exhibits a transition in the asymptotic long-term dynamics as a function of coupling strength $\Gamma$:
For negligible couplings $\Gamma$, condition~\eqref{EQ:boundstatecondition} is not fulfilled, and the system thermalizes with the reservoir temperature.
For finite but still small coupling strengths disobeying~\eqref{EQ:boundstatecondition}, the system alone approaches a non-thermal (with regard to the system Hamiltonian $H_S$) steady-state~\cite{hilt2011a,timofeev2022a,burke2024a}.
And for large coupling strengths obeying~\eqref{EQ:boundstatecondition}, no steady state is reached at all, with the stationary contribution becoming more and more suppressed with increasing coupling strength $\Gamma$, even at finite temperatures.
While the detailed analysis of this is provided in Appendix~\ref{APP:exactsolution}, we also provide a different perspective with which this behaviour can be understood using RCs below.

\section{Parallel RC mapping}\label{SEC:rcmapping}

Reaction-coordinate mappings are often used to convert star representations (left panel of Fig.~\ref{FIG:rcmappingspecdens}) to chain-star representations or even chain representations by repeated applications~\cite{martinazzo2011a,woods2014a,landi2022a}.
We discuss the derivation for the standard RC mapping iteration in App.~\ref{APP:rcmapping:single}.
Particularly the Rubin spectral function~\eqref{EQ:sfrubin} is up to a constant inert under the standard mapping (see App.~\ref{APP:rcmapping:rubin}), 
such that even for repeated application, a weak-coupling treatment may not apply. 
Contrary to the standard approach, we therefore partition the region of support of $\Gamma(\omega)$ into $N$ disjoint intervals ${\cal I}_i$ and perform the RC mapping for each interval (see inset in the right panel of Fig.~\ref{FIG:rcmappingspecdens}), which yields
\begin{align}\label{EQ:model3}
H &= H_S + \sum_{i=1}^N \Omega_i \left[B_i^\dagger + \frac{\lambda_i}{\Omega_i} (a+a^\dagger)\right]\left[B_i + \frac{\lambda_i}{\Omega_i} (a+a^\dagger)\right]\nn
&\qquad+ \sum_{i,q}\omega_{iq} B_{iq}^\dagger B_{iq} + \sum_{i,q} (B_i+B_i^\dagger)H_{iq}(B_{iq}+B_{iq}^\dagger)\,.
\end{align}
Here, the first line denotes the supersystem composed of the original system coupled to the RCs with energies $\Omega_i$ via coupling strengths $\lambda_i$.
Each bosonic RC ($B_i,B_i^\dagger$) is coupled to its residual sub-reservoir via coupling amplitudes $H_{iq}$.
We stress that the mapping allows to treat the interaction Hamiltonian as part of the supersystem and thus provides a physical interpretation of the RCs~\cite{iles_smith2014a,iles_smith2016a,newman2017a} as it explicitly demands
$H_I = \sum_k (a+a^\dagger) (h_k b_k + h_k^* b_k^\dagger) = \sum_i \lambda_i (a+a^\dagger)(B_i+B_i^\dagger)$, which is less apparent in other approaches employing parallel mappings~\cite{garraway1997a,huh2014a,pleasance2020a}.
The amplitudes allow to define a residual spectral function $\Gamma_i(\omega)=2\pi\sum_q \abs{H_{iq}}^2 \delta(\omega-\omega_{iq})$.
In the continuum limit, these parameters can be obtained from the original spectral function via
\begin{align}\label{EQ:rcmapping2}
\Omega_i^2 &= \frac{\int_{{\cal I}_i} d\omega \omega \Gamma(\omega)}{\int_{{\cal I}_i} d\omega \frac{\Gamma(\omega)}{\omega}}\,,\nn
\lambda_i^2 &= \frac{1}{2\pi\Omega_i}\int_{{\cal I}_i} d\omega \omega \Gamma(\omega)\,,\nn
\Gamma_i(\omega) &\stackrel{\omega\in{\cal I}_i}{=} \frac{4 \lambda_i^2 \Gamma(\omega)}{\Gamma^2(\omega) + \left(\frac{1}{\pi} {\cal P} \int\limits_{{\cal I}_i \cup \bar{\cal I}_i} d\omega' \frac{\Gamma(\omega')}{\omega'-\omega}\right)^2}\,.
\end{align}
Here, $\bar{\cal I}_i$ denotes the corresponding negative interval (${\cal I}_i=[\omega_a,\omega_b]$ implies $\bar{\cal I}_i = [-\omega_b,-\omega_a]$), 
and in the principal value integral of the last equation $\Gamma(\omega)$ has to be continued as an odd function to the complete real axis $\Gamma(-\omega)=-\Gamma(\omega)$.
In Fig.~\ref{FIG:rcmappingspecdens}, the right panel shows the result of the mapping applied to the left panel spectral function, where one can see that the residual couplings are significantly smaller and
that the saturation limit of $\Gamma_i(\omega) \lesssim \frac{2}{\pi}\Delta\omega$ for small discretization widths (see App.~\ref{APP:rcmapping:weakvariation}) is already well respected.
We find it more convenient to define the interval width via demanding that 
\begin{align}\label{EQ:discretization}
2\pi\lambda_i^2\Omega_i = \int_{{\cal I}_i} d\omega \omega \Gamma(\omega) = \frac{\int d\omega\omega\Gamma(\omega)}{N}\,,
\end{align}
as this not only allows to partition infinite regions into finitely many intervals but also implies that the quantities $\lambda_i^2 \Omega_i$ are all identical.

In the particular case that the system is just harmonic $H_S=\Omega a^\dagger a$, the squared supersystem excitation energies $\epsilon_q^2$ are given by the eigenvalues of the $(N+1)\times(N+1)$ matrix
(see App.~\ref{APP:rcmapping:excitations})
\begin{align}\label{EQ:excitationmatrix}
M &=\left(\begin{array}{cccc}
     \Omega^2+\sum_i\frac{4\Omega\lambda_i^2}{\Omega_i} &2\lambda_1\sqrt{\Omega\Omega_1}& 2\lambda_2\sqrt{\Omega\Omega_2} & \cdots  \\
   2\lambda_1\sqrt{\Omega\Omega_1}&\Omega_1^2 &0&\cdots\\
   2\lambda_2\sqrt{\Omega\Omega_2}&0&\Omega_2^2&\cdots\\
    \vdots& \vdots& \vdots&\ddots
\end{array}\right)\,.
\end{align}
From the arrowhead structure and positive definiteness we can conclude that the $N+1$ eigenvalues obey
$0 \le \epsilon_1^2 \le \Omega_1^2 \le \epsilon_2^2 \le \ldots \le \Omega_N^2 \le \epsilon_{N+1}^2$, which pins the first $N$ eigenvalues between the RC energies inside the band as outlined in App.~\ref{APP:arrowhead:all}.
For the largest eigenvalue, we obtain the simple lower bound 
\begin{align}\label{EQ:boundlooselower}
\epsilon_{N+1}^2 \ge \Omega^2 + \frac{2\Omega}{\pi} \int d\omega \frac{\Gamma(\omega)}{\omega}\,.
\end{align}
We explain this and also further upper and tighter lower bounds in App.~\ref{APP:arrowhead:largest}.
As $\epsilon_{N+1}$ definitely leaves the band (i.e., for our example~\eqref{EQ:sfrubin} when $\epsilon_{N+1} \ge \omega_c$) for sufficiently large coupling strengths, this shows that regardless of the actual shape of $\Gamma(\omega)$, 
an isolated level (the BS) will exist beyond a critical coupling strength -- trivially so, when already $\Omega$ is outside the band.

\begin{figure}[h]
\includegraphics[width=0.45\textwidth]{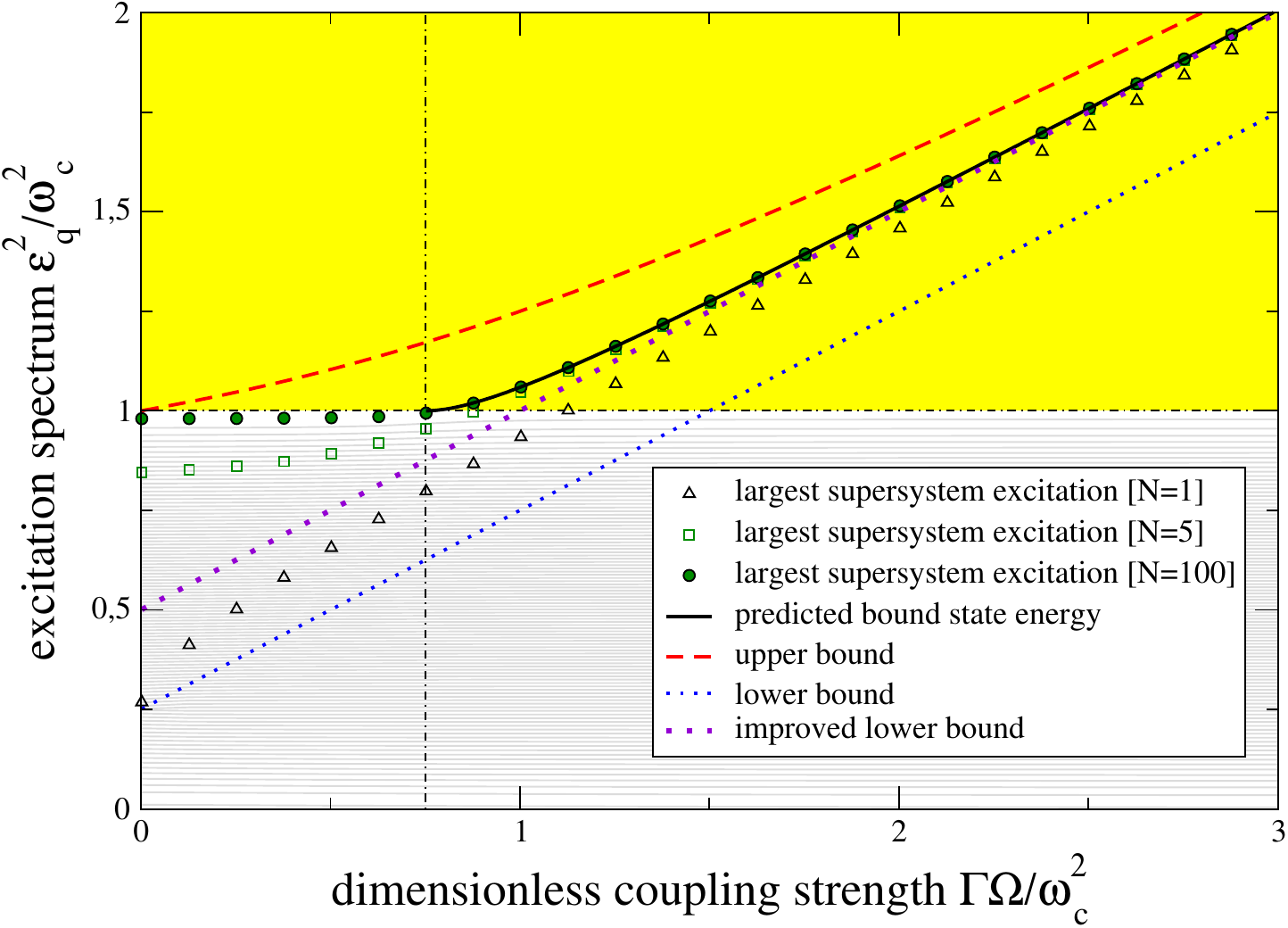}
\caption{\label{FIG:boundstatewithrc}
Plot of the squared excitation energies versus dimensionless coupling strength (symbols). 
Beyond a coupling strength defined by~\eqref{EQ:boundstatecondition} (vertical dash-dotted line), the highest supersystem mode (green circles) enters the band gap (yellow region) and forms a BS -- matching the exact prediction~\eqref{EQ:prediction_boundstate} (solid black).
All other supersystem eigenvalues remain within the band (thin gray for $N=100$). 
Bounds apply only to the largest eigenvalue.
Other parameters: $\Omega=\omega_c/2$.
}
\end{figure}

In Fig.~\ref{FIG:boundstatewithrc} we plot the squared excitation energies vs. the coupling strength and find these properties well reproduced.
One can see that beyond the critical coupling strength~\eqref{EQ:boundstatecondition}, the largest eigenvalue departs from the others and enters the band gap (yellow region).
A sufficiently fine discretization provided, the onset of the BS formation agrees with the exact solution~\eqref{EQ:prediction_boundstate}.
Then, at least a partial secular treatment may be applied (see App.~\ref{APP:perturbative}), which immediately implies that the BS is stable as also the
residual reservoir supports the same band gap, see Fig.~\ref{FIG:rcmappingspecdens} right panel.

\section{Extensions}\label{SEC:extensions}

\subsection{Multiple band gaps}

\begin{figure}
\includegraphics[width=0.45\textwidth]{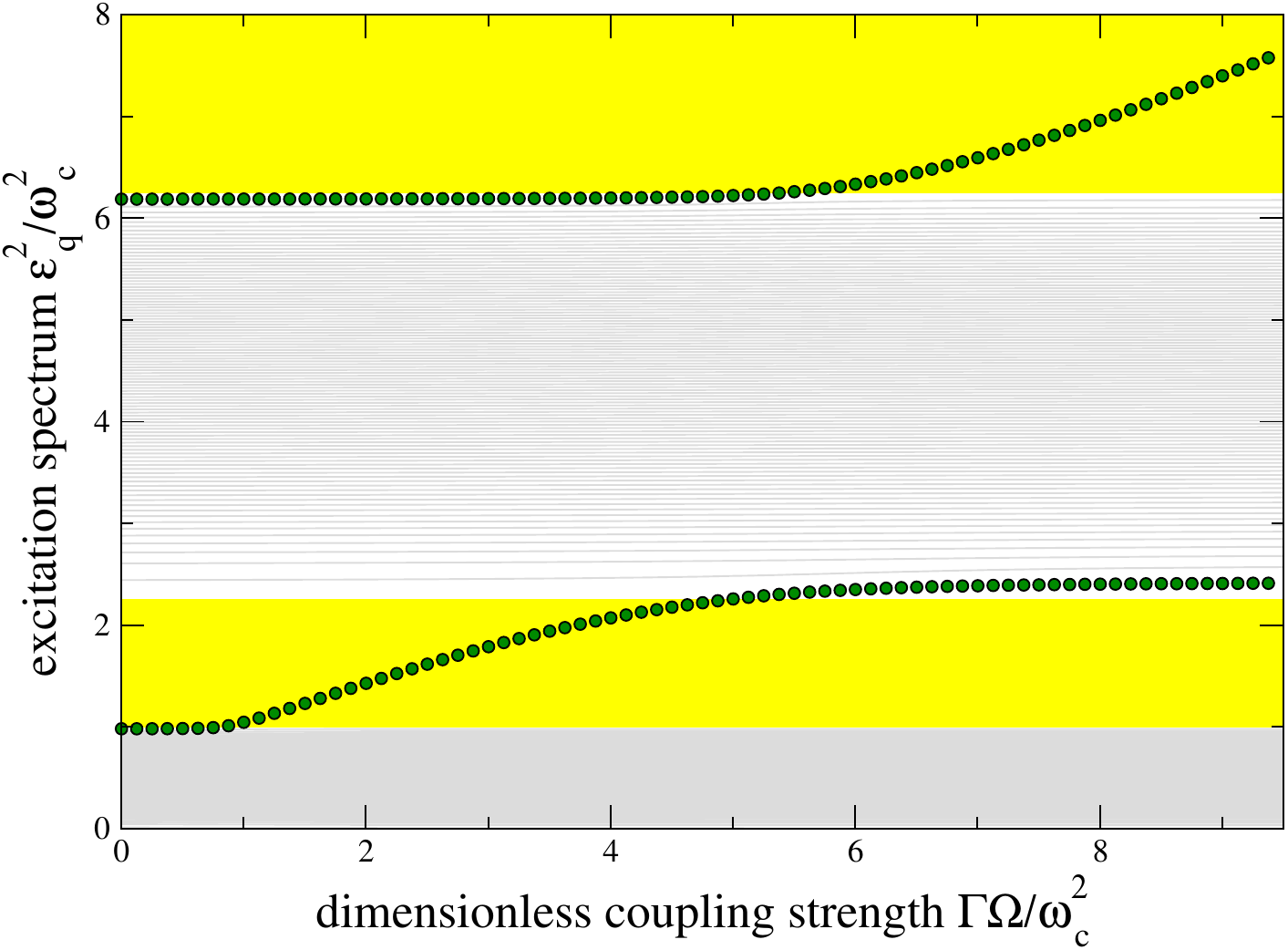}
\caption{\label{FIG:boundstate2bands}
Analogous to Fig.~\ref{FIG:boundstatewithrc}, but with two bands generated by a superposing~\eqref{EQ:sfrubin} twice by using the spectral function $\Gamma(\omega) + \Gamma(\omega-\frac{3}{2}\omega_c)$.
The system energy was inside the first band $\Omega=\omega_c/2$. For each band, we used 100 RCs, and the eigenvalues that may potentially form a BS are represented by green circles.
}
\end{figure}

For a spectral function with multiple band gaps, that has e.g. support in multiple finite intervals, the RC energies will -- sufficiently fine discretization provided -- cluster inside the bands. 
Then, by cutting the first line and first row of the excitation matrix~\eqref{EQ:excitationmatrix}, the Poincar{\'e} separation theorem would still bound the eigenvalues between the squared RC energies. 
Particularly, there would be a single excitation matrix eigenvalue between the largest eigenvalue of band $i$ and the smallest eigenvalue of band $i+1$.
This implies that every band gap can support one BS at most.
While for an infinitely large band gap, the energy of the BS can grow indefinitely -- for our example~\eqref{EQ:sfrubin} it scales as $\epsilon_{N+1}^2 \approx \frac{\Gamma\Omega}{2} + \Omega^2+\frac{\omega_c^2}{4}$ for large couplings -- this is different for the eigenvalues of constrained band gaps.
From the trace of the excitation matrix we may conclude that at very strong couplings, the largest eigenvalue leaves into the largest band gap, it is thus not necessarily the case that each finite-width band gap hosts a BS at a fixed coupling strength.
The numerical example that we consider in Fig.~\ref{FIG:boundstate2bands} suggests that there are regimes where only one BS exists at a certain coupling strength.
As before, we find that at very weak coupling strengths (and presupposing that $\Omega$ is inside one of the bands), there exists no BS.
Beyond a critical coupling, a BS forms in the first band gap, further increases in energy and then leaves the band gap again, loosing its immunity.
Then, it forms again in the unconstrained band gap (top).

\subsection{Interactions}

One may wonder about the fate of the BS in presence of integrability-breaking terms. 
For example, if the system Hamiltonian contains anharmonic interactions, e.g. $H_S = \Omega a^\dagger a + U (a+a^\dagger)^4$, an exact solution is not available and one can also not determine the supersystem excitation energies via the eigenvalues of~\eqref{EQ:excitationmatrix}.
Nevertheless, one may adress what happens in the weakly interacting limit $U \ll \Gamma,\Omega,\omega_c$.
Then, we may use a single RC ($N=1$) and employ a perturbative treatment of the system-reservoir ($\Gamma$) and the anharmonic ($U$) interactions on equal footing.
The interaction picture is then obtained with respect to the quadratic part of the Hamiltonian, and we obtain the same dissipator as in the absence of interactions, technically analogous
to the derivation of a local master equation~\cite{schaller2022a,landi2022a}, see also App.~\ref{APP:perturbative}.
The anharmonic interaction only remains in the unitary part, which however leads to drastic differences in the dynamics.

Under a secular treatment of the supersystem, we would for $\Omega=\omega_c/2$ obtain the Lindblad-Gorini-Kossakowski-Sudarshan (LGKS) master equation
\begin{align}
\dot\rho &= -\ii \left[\epsilon_1 c_1^\dagger c_1 + \epsilon_2 c_2^\dagger c_2 + \frac{U\omega_c}{\Gamma} (c_2+c_2^\dagger+c_1+c_1^\dagger)^4, \rho\right]\nn
&\qquad+\sqrt{\frac{\Gamma}{\omega_c}} \Gamma_1(\epsilon_1)[1+n(\epsilon_1)] \left[c_1 \rho c_1^\dagger - \frac{1}{2} \left\{c_1^\dagger c_1, \rho\right\}\right]\nn
&\qquad+\sqrt{\frac{\Gamma}{\omega_c}} \Gamma_1(\epsilon_1)n(\epsilon_1) \left[c_1^\dagger \rho c_1 - \frac{1}{2} \left\{c_1 c_1^\dagger, \rho\right\}\right]\,,
\end{align}
where in the strong-coupling regime $\epsilon_2\approx \sqrt{\Gamma\omega_c}/2$ and $\Gamma_1(\epsilon_1)\approx \epsilon_1 \approx \frac{\omega_c^{3/2}}{2\Gamma^{1/2}}$, such that the supersystem dissipator is actually not dependent on $\Gamma$ -- highlighting the usefulness of the RC approach.
Here, we have inserted the strong-coupling regime for $U(a+a^\dagger)^4$ from Eq.~\eqref{EQ:bogoliubov1}.
Obviously, any BS variable depending solely on $c_2$ and $c_2^\dagger$ is not directly affected by the last two lines (i.e., the dissipator).
This means that to decay, the BS has to be rotated first into the vulnerable sector, which implies a lower bound on its lifetime
\begin{align}
\tau_b \gtrsim \ord\left\{\frac{\Gamma}{U\omega_c}\right\}\,.
\end{align}
This reveals that although interactions spoil the immortality of the BS, its stability can be increased by large system-reservoir coupling strength, small reservoir bandwidth and of course small anharmonicity.
A similar picture arises if many RCs are used, though the dissipator may not admit a full secular treatment, see App.~\ref{APP:rcmapping:excitations}.

\section{Summary and Outlook}\label{SEC:summary}

Our study demonstrates that parallel RCs can be used to model strong-coupling phenomena as by using sufficiently many RCs the residual coupling becomes smaller and smaller.
When applied to a harmonic model, the RC picture reveals the existence of BSs for any gapped spectral function once the coupling strength is large enough -- a loose lower bound
may be used for that.
The price one has to pay is the computational effort required to model the RCs explicitly (moderate for non-interacting models).
We found that weak interactions can be studied perturbatively, which allows to investigate the onset of ergodicity in originally integrable models~\cite{rigol2007a,essler2016a,wang2025a,wang2025b}.

We also find that the RC energies and the resulting excitation energies of the supersystem are not necessarily far apart from each other, such that a full secular LGKS treatment of the supersystem may not apply. 
Nevertheless, one may solve it using Redfield approaches~\cite{redfield1965a,hartmann2020a} or generalized LGKS treatments~\cite{wichterich2007a,schultz2009a,kirsanskas2018a,kleinherbers2020a,nathan2020a,trushechkin2021a}.
However, the ultra-strong coupling limit can already be well treated with a single RC and a secular LGKS treatment.

In addition, we note that the approach of partitioning the spectral functions support can be equally applied to fermions in a straightforward way. 
In this case, one has to use the particle mapping~\cite{woods2014a,nazir2019a}, and due to the finite Hilbert space dimension an explicit modelling
is simpler than in the bosonic case.
As an outlook, also the multi-reservoir case~\cite{stefanucci2007a,khosravi2009a,jussiau2019a} can be generalized beyond the single RC per reservoir limit, and the same applies to driven systems~\cite{restrepo2019a,litzba2026a}.

\section{Acknowledgments}

G.S. and F.Q. acknowledge helpful discussions with P. Strasberg and support by the CRC 1242 (DFG-project 278162697).
G.-Y. L. has been supported by the HZDR summer student program.

\bibliographystyle{unsrt}
\bibliography{references}

\begin{thebibliography}{10}

\bibitem{breuer2007}
Heinz-Peter Breuer and Francesco Petruccione.
\newblock {\em The Theory of Open Quantum Systems}.
\newblock Oxford University Press, 01 2007.

\bibitem{wolf2008a}
M.~M. Wolf, J.~Eisert, T.~S. Cubitt, and J.~I. Cirac.
\newblock Assessing non-{M}arkovian quantum dynamics.
\newblock {\em Phys. Rev. Lett.}, 101:150402, Oct 2008.

\bibitem{breuer2009c}
Heinz-Peter Breuer, Elsi-Mari Laine, and Jyrki Piilo.
\newblock Measure for the degree of non-{M}arkovian behavior of quantum
  processes in open systems.
\newblock {\em Phys. Rev. Lett.}, 103:210401, Nov 2009.

\bibitem{rivas2010a}
\'Angel Rivas, Susana~F. Huelga, and Martin~B. Plenio.
\newblock Entanglement and non-{M}arkovianity of quantum evolutions.
\newblock {\em Phys. Rev. Lett.}, 105:050403, Jul 2010.

\bibitem{breuer2012a}
Heinz-Peter Breuer.
\newblock Foundations and measures of quantum non-{M}arkovianity.
\newblock {\em Journal of Physics B: Atomic, Molecular and Optical Physics},
  45(15):154001, jul 2012.

\bibitem{unruh1995a}
W.~G. Unruh.
\newblock Maintaining coherence in quantum computers.
\newblock {\em Physical Review A}, 51:992--997, 1995.

\bibitem{reina2002a}
John~H. Reina, Luis Quiroga, and Neil~F. Johnson.
\newblock Decoherence of quantum registers.
\newblock {\em Physical Review A}, 65:032326, 2002.

\bibitem{divincenzo2000a}
David~P. DiVincenzo.
\newblock The physical implementation of quantum computation.
\newblock {\em Fortschritte der Physik}, 48:771--783, 2000.

\bibitem{nielsen2000}
Michael~A. Nielsen and Isaac~L. Chuang.
\newblock {\em {Quantum Computation and Quantum Information}}.
\newblock Cambridge University Press, Cambridge, 2000.

\bibitem{mahan1990}
Gerald~D. Mahan.
\newblock {\em Many-particle physics}.
\newblock Springer, New York, 2nd edition, 1990.

\bibitem{marinica2008a}
D.~C. Marinica, A.~G. Borisov, and S.~V. Shabanov.
\newblock Bound states in the continuum in photonics.
\newblock {\em Phys. Rev. Lett.}, 100:183902, May 2008.

\bibitem{hsu2016a}
Chia~Wei Hsu, Bo~Zhen, A.~Douglas Stone, John~D. Joannopoulos, and Marin
  Solja\v{c}i\'c.
\newblock Bound states in the continuum.
\newblock {\em Nature Reviews Materials}, 1(9):16048, 2016.

\bibitem{longhi2007a}
S.~Longhi.
\newblock Bound states in the continuum in a single-level {F}ano-{A}nderson
  model.
\newblock {\em The European Physical Journal B}, 57:45--51, 2007.

\bibitem{john1990a}
Sajeev John and Jian Wang.
\newblock Quantum electrodynamics near a photonic band gap: Photon bound states
  and dressed atoms.
\newblock {\em Phys. Rev. Lett.}, 64:2418--2421, May 1990.

\bibitem{kofman1994a}
A.G. Kofman, G.~Kurizki, and B.~Sherman.
\newblock Spontaneous and induced atomic decay in photonic band structures.
\newblock {\em Journal of Modern Optics}, 41(2):353--384, 1994.

\bibitem{angelakis2004a}
D.~G. Angelakis, P.~L. Knight, and E.~Paspalakis.
\newblock Photonic crystals and inhibition of spontaneous emission: an
  introduction.
\newblock {\em Contemporary Physics}, 45(4):303--318, 2004.

\bibitem{chang2018a}
D.~E. Chang, J.~S. Douglas, A.~Gonz\'alez-Tudela, C.-L. Hung, and H.~J. Kimble.
\newblock Colloquium: Quantum matter built from nanoscopic lattices of atoms
  and photons.
\newblock {\em Rev. Mod. Phys.}, 90:031002, Aug 2018.

\bibitem{dhar2006a}
Abhishek Dhar and Diptiman Sen.
\newblock Nonequilibrium {G}reen's function formalism and the problem of bound
  states.
\newblock {\em Phys. Rev. B}, 73:085119, Feb 2006.

\bibitem{stefanucci2007a}
Gianluca Stefanucci.
\newblock Bound states in ab initio approaches to quantum transport: A
  time-dependent formulation.
\newblock {\em Phys. Rev. B}, 75:195115, May 2007.

\bibitem{jussiau2019a}
\'Etienne Jussiau, Masahiro Hasegawa, and Robert~S. Whitney.
\newblock Signature of the transition to a bound state in thermoelectric
  quantum transport.
\newblock {\em Phys. Rev. B}, 100:115411, Sep 2019.

\bibitem{plotnik2011a}
Yonatan Plotnik, Or~Peleg, Felix Dreisow, Matthias Heinrich, Stefan Nolte,
  Alexander Szameit, and Mordechai Segev.
\newblock Experimental observation of optical bound states in the continuum.
\newblock {\em Phys. Rev. Lett.}, 107:183901, Oct 2011.

\bibitem{amrani2021a}
Madiha Amrani, Ilyasse Quotane, Cecile Ghouila-Houri, El~Houssaine El~Boudouti,
  Leonid Krutyansky, Bogdan Piwakowski, Philippe Pernod, Abdelkrim Talbi, and
  Bahram Djafari-Rouhani.
\newblock Experimental evidence of the existence of bound states in the
  continuum and {F}ano resonances in solid-liquid layered media.
\newblock {\em Phys. Rev. Appl.}, 15:054046, May 2021.

\bibitem{martinazzo2011a}
R.~Martinazzo, B.~Vacchini, K.~H. Hughes, and I.~Burghardt.
\newblock Communication: Universal {M}arkovian reduction of {B}rownian particle
  dynamics.
\newblock {\em The Journal of Chemical Physics}, 134(1):011101, 01 2011.

\bibitem{woods2014a}
M.~P. Woods, R.~Groux, A.~W. Chin, S.~F. Huelga, and M.~B. Plenio.
\newblock Mappings of open quantum systems onto chain representations and
  {M}arkovian embeddings.
\newblock {\em Journal of Mathematical Physics}, 55:032101, 2014.

\bibitem{strasberg2016a}
Philipp Strasberg, Gernot Schaller, Neill Lambert, and Tobias Brandes.
\newblock Nonequilibrium thermodynamics in the strong coupling and
  non-{M}arkovian regime based on a reaction coordinate mapping.
\newblock {\em New Journal of Physics}, 18:073007, 2016.

\bibitem{nazir2019a}
A.~Nazir and G.~Schaller.
\newblock The reaction coordinate mapping in quantum thermodynamics.
\newblock In F.~Binder, L.~A. Correa, C.~Gogolin, J.~Anders, and G.~Adesso,
  editors, {\em Thermodynamics in the quantum regime -- Recent progress and
  outlook}, Fundamental Theories of Physics, page 551. Springer, Cham, 2019.

\bibitem{correa2019a}
Luis~A. Correa, Buqing Xu, and Benjamin Morris~Gerardo Adesso.
\newblock Pushing the limits of the reaction-coordinate mapping.
\newblock {\em Journal of Chemical Physics}, 151:094107, 2019.

\bibitem{anto_sztrikacs2021a}
Nicholas Anto-Sztrikacs and Dvira Segal.
\newblock Strong coupling effects in quantum thermal transport with the
  reaction coordinate method.
\newblock {\em New Journal of Physics}, 23(6):063036, jun 2021.

\bibitem{leggett1987a}
A.~J. Leggett, S.~Chakravarty, A.~T. Dorsey, Matthew P.~A. Fisher, Anupam Garg,
  and W.~Zwerger.
\newblock Dynamics of the dissipative two-state system.
\newblock {\em Rev. Mod. Phys.}, 59:1--85, Jan 1987.

\bibitem{schaller2009a}
Gernot Schaller, Philipp Zedler, and Tobias Brandes.
\newblock Systematic perturbation theory for dynamical coarse-graining.
\newblock {\em Phys. Rev. A}, 79:032110, Mar 2009.

\bibitem{zhang2012a}
Wei-Min Zhang, Ping-Yuan Lo, Heng-Na Xiong, Matisse Wei-Yuan Tu, and Franco
  Nori.
\newblock General non-{M}arkovian dynamics of open quantum systems.
\newblock {\em Phys. Rev. Lett.}, 109:170402, Oct 2012.

\bibitem{topp2015a}
Gabriel~E. Topp, Tobias Brandes, and Gernot Schaller.
\newblock Steady-state thermodynamics of non-interacting transport beyond weak
  coupling.
\newblock {\em Europhysics Letters}, 110:67003, 2015.

\bibitem{rubin1963a}
Robert~J. Rubin.
\newblock Momentum autocorrelation functions and energy transport in harmonic
  crystals containing isotopic defects.
\newblock {\em Phys. Rev.}, 131:964--989, Aug 1963.

\bibitem{weiss1993}
U.~Weiss.
\newblock {\em Quantum Dissipative Systems}, volume~2 of {\em Series of Modern
  Condensed Matter Physics}.
\newblock World Scientific, Singapore, 1993.

\bibitem{hilt2011a}
Stefanie Hilt, Benedikt Thomas, and Eric Lutz.
\newblock Hamiltonian of mean force for damped quantum systems.
\newblock {\em Phys. Rev. E}, 84:031110, Sep 2011.

\bibitem{timofeev2022a}
G.~M. Timofeev and A.~S. Trushechkin.
\newblock Hamiltonian of mean force in the weak-coupling and high-temperature
  approximations and refined quantum master equations.
\newblock {\em International Journal of Modern Physics A}, 37(20n21):2243021,
  2022.

\bibitem{burke2024a}
Phillip~C. Burke, Goran Nakerst, and Masudul Haque.
\newblock Structure of the {H}amiltonian of mean force.
\newblock {\em Phys. Rev. E}, 110:014111, Jul 2024.

\bibitem{landi2022a}
Gabriel~T. Landi, Dario Poletti, and Gernot Schaller.
\newblock Nonequilibrium boundary-driven quantum systems: Models, methods, and
  properties.
\newblock {\em Rev. Mod. Phys.}, 94:045006, Dec 2022.

\bibitem{iles_smith2014a}
Jake Iles-Smith, Neill Lambert, and Ahsan Nazir.
\newblock Environmental dynamics, correlations, and the emergence of
  noncanonical equilibrium states in open quantum systems.
\newblock {\em Phys. Rev. A}, 90:032114, Sep 2014.

\bibitem{iles_smith2016a}
Jake Iles-Smith, Arend~G. Dijkstra, Neill Lambert, and Ahsan Nazir.
\newblock Energy transfer in structured and unstructured environments: Master
  equations beyond the {B}orn-{M}arkov approximations.
\newblock {\em The Journal of Chemical Physics}, 144(4):044110, 01 2016.

\bibitem{newman2017a}
David Newman, Florian Mintert, and Ahsan Nazir.
\newblock Performance of a quantum heat engine at strong reservoir coupling.
\newblock {\em Physical Review E}, 95:032139, 2017.

\bibitem{garraway1997a}
B.~M. Garraway.
\newblock Nonperturbative decay of an atomic system in a cavity.
\newblock {\em Phys. Rev. A}, 55:2290--2303, Mar 1997.

\bibitem{huh2014a}
Joonsuk Huh, Sarah Mostame, Takatoshi Fujita, Man-Hong Yung, and Alán
  Aspuru-Guzik.
\newblock Linear-algebraic bath transformation for simulating complex open
  quantum systems.
\newblock {\em New Journal of Physics}, 16(12):123008, dec 2014.

\bibitem{pleasance2020a}
Graeme Pleasance, Barry~M. Garraway, and Francesco Petruccione.
\newblock Generalized theory of pseudomodes for exact descriptions of
  non-{M}arkovian quantum processes.
\newblock {\em Phys. Rev. Research}, 2:043058, Oct 2020.

\bibitem{schaller2022a}
G.~Schaller, F.~Queisser, N.~Szpak, J.~K\"onig, and R.~Sch\"utzhold.
\newblock Environment-induced decay dynamics of antiferromagnetic order in
  {M}ott-{H}ubbard systems.
\newblock {\em Phys. Rev. B}, 105:115139, Mar 2022.

\bibitem{rigol2007a}
Marcos Rigol, Vanja Dunjko, Vladimir Yurovsky, and Maxim Olshanii.
\newblock Relaxation in a completely integrable many-body quantum system: An ab
  initio study of the dynamics of the highly excited states of 1d lattice
  hard-core bosons.
\newblock {\em Phys. Rev. Lett.}, 98:050405, Feb 2007.

\bibitem{essler2016a}
Fabian H~L Essler and Maurizio Fagotti.
\newblock Quench dynamics and relaxation in isolated integrable quantum spin
  chains.
\newblock {\em Journal of Statistical Mechanics: Theory and Experiment},
  2016(6):064002, jun 2016.

\bibitem{wang2025a}
Jiaozi Wang, Merlin F\"ullgraf, and Jochen Gemmer.
\newblock Estimate of equilibration times of quantum correlation functions in
  the thermodynamic limit based on {L}anczos coefficients.
\newblock {\em Phys. Rev. Lett.}, 135:010403, Jul 2025.

\bibitem{wang2025b}
Jiaozi Wang and Philipp Strasberg.
\newblock Decoherence of histories: Chaotic versus integrable systems.
\newblock {\em Phys. Rev. Lett.}, 134:220401, Jun 2025.

\bibitem{redfield1965a}
A.~G. Redfield.
\newblock {\em Advances in Magnetic and Optical Resonance}, chapter The Theory
  of Relaxation Processes, pages 1--32.
\newblock Advances in Magnetic and Optical Resonance. Academic Press, New York,
  1965.

\bibitem{hartmann2020a}
Richard Hartmann and Walter~T. Strunz.
\newblock Accuracy assessment of perturbative master equations: Embracing
  nonpositivity.
\newblock {\em Phys. Rev. A}, 101:012103, Jan 2020.

\bibitem{wichterich2007a}
Hannu Wichterich, Markus~J. Henrich, Heinz-Peter Breuer, Jochen Gemmer, and
  Mathias Michel.
\newblock Modeling heat transport through completely positive maps.
\newblock {\em Physical Review E}, 76(3):031115, 2007.

\bibitem{schultz2009a}
M.~G. Schultz and F.~von Oppen.
\newblock Quantum transport through nanostructures in the singular-coupling
  limit.
\newblock {\em Physical Review B}, 80:033302, 2009.

\bibitem{kirsanskas2018a}
Gediminas Kir\ifmmode~\check{s}\else \v{s}\fi{}anskas, Martin Francki\'e, and
  Andreas Wacker.
\newblock Phenomenological position and energy resolving {L}indblad approach to
  quantum kinetics.
\newblock {\em Phys. Rev. B}, 97:035432, Jan 2018.

\bibitem{kleinherbers2020a}
Eric Kleinherbers, Nikodem Szpak, J\"urgen K\"onig, and Ralf Sch\"utzhold.
\newblock Relaxation dynamics in a {H}ubbard dimer coupled to fermionic baths:
  Phenomenological description and its microscopic foundation.
\newblock {\em Phys. Rev. B}, 101:125131, Mar 2020.

\bibitem{nathan2020a}
Frederik Nathan and Mark~S. Rudner.
\newblock Universal {L}indblad equation for open quantum systems.
\newblock {\em Phys. Rev. B}, 102:115109, Sep 2020.

\bibitem{trushechkin2021a}
Anton Trushechkin.
\newblock Unified {G}orini-{K}ossakowski-{L}indblad-{S}udarshan quantum master
  equation beyond the secular approximation.
\newblock {\em Phys. Rev. A}, 103:062226, Jun 2021.

\bibitem{khosravi2009a}
E.~Khosravi, G.~Stefanucci, S.~Kurth, and E.K.U. Gross.
\newblock Bound states in time-dependent quantum transport: oscillations and
  memory effects in current and density.
\newblock {\em Phys. Chem. Chem. Phys.}, 11:4535--4538, 2009.

\bibitem{restrepo2019a}
Sebastian Restrepo, Sina B\"ohling, Javier Cerrillo, and Gernot Schaller.
\newblock Electron pumping in the strong coupling and non-{M}arkovian regime: A
  reaction coordinate mapping approach.
\newblock {\em Phys. Rev. B}, 100:035109, Jul 2019.

\bibitem{litzba2026a}
Lukas Litzba, Gernot Schaller, Jürgen König, and Nikodem Szpak.
\newblock Coupling-energy driven pumping through quantum dots: the role of
  coherences, 2026.

\bibitem{varga2001a}
Richard~S. Varga.
\newblock Gerschgorin disks, {B}rauer ovals of {C}assini (a vindication), and
  {B}rualdi sets.
\newblock {\em Information}, 4(2):171--178, 2001.

\bibitem{farina2019a}
Donato Farina and Vittorio Giovannetti.
\newblock Open-quantum-system dynamics: Recovering positivity of the {R}edfield
  equation via the partial secular approximation.
\newblock {\em Phys. Rev. A}, 100:012107, Jul 2019.

\bibitem{cattaneo2019a}
Marco Cattaneo, Gian~Luca Giorgi, Sabrina Maniscalco, and Roberta Zambrini.
\newblock Local versus global master equation with common and separate baths:
  superiority of the global approach in partial secular approximation.
\newblock {\em New Journal of Physics}, 21:113045, 2019.

\end{thebibliography}

\appendix

\section{Exact solution}\label{APP:exactsolution}

\subsection{Heisenberg picture}\label{APP:exactsolution:heisenberg}

Setting up the Heisenberg equations of motion, we obtain from~\eqref{EQ:model1} the equations for the operators in the Heisenberg picture (bold symbols)
\begin{align}
\f{\dot x} &= \Omega \f{p}\,,\nn
\f{\dot p} &= -\left[\Omega+4\sum_k \frac{\abs{h_k}^2}{\omega_k}\right]\f{x}
-2 \sum_k (h_k \f{b_k} + h_k^* \f{b_k^\dagger})\,,\nn
\f{\dot b_k} &= -\ii h_k^* \f{x} - \ii \omega_k \f{b_k}\,,\nn
\f{\dot b_k^\dagger} &= +\ii h_k \f{x} + \ii \omega_k \f{b_k^\dagger}\,,
\end{align}
which can be solved by Laplace-transforming the operators
according to e.g. $x(z) = \int_0^\infty \f{x} e^{-z t} dt$ and analogously for the others
(we omit writing the Laplace transform of hermitian conjugate operators explicitly).
This yields an algebraic system ($\bar b_k(z)$ denotes the Laplace transform of $\f{b_k^\dagger}$)
\begin{align}
z x(z) - x &= \Omega p(z)\,,\nn
z p(z) - p &= -\left[\Omega+4\sum_k \frac{\abs{h_k}^2}{\omega_k}\right]x(z)\nn
&\qquad-2 \sum_k (h_k b_k(z) + h_k^* \bar b_k(z))\,,\nn
z b_k(z) - b_k &= -\ii h_k^* x(z) - \ii \omega_k b_k(z)\,.
\end{align}
Eliminating first the reservoir modes, solving then the system equations and eventually inserting the solution in the reservoir modes
allows to represent the Laplace-transformed operators in terms of their initial (Schr\"odinger picture) operators
\begin{align}\label{EQ:solution_laplace}
x(z) &= \frac{z x + \Omega  p}{z^2+\Omega f(z)}\\
&\qquad- \frac{2\Omega}{z^2+\Omega f(z)}
\sum_k \left(\frac{h_k  b_k}{z+\ii\omega_k}+\frac{h_k^*  b_k^\dagger}{z-\ii\omega_k}\right)\,,\nn
p(z) &= \frac{z  p - f(z)  x}{z^2+\Omega f(z)}\nn
&\qquad - \frac{2z}{z^2+\Omega f(z)}
\sum_k \left(\frac{h_k  b_k}{z+\ii\omega_k}+\frac{h_k^*  b_k^\dagger}{z-\ii\omega_k}\right)\,,\nn
b_k(z) &= \frac{1}{z+\ii\omega_k} b_k - \frac{\ii h_k^*}{z+\ii\omega_k} \Big[
 \frac{z x + \Omega p}{z^2+\Omega f(z)}\nn
 &\qquad - \frac{2\Omega}{z^2+\Omega f(z)}\sum_q \left(\frac{h_q b_q}{z+\ii\omega_q}+\frac{h_q^* b_q^\dagger}{z-\ii\omega_q}\right)\Big]\,.\nonumber
\end{align}
Here, the function $f(z)$ is given by
\begin{align}
f(z)=\Omega + \frac{2}{\pi} \int_0^\infty \frac{\Gamma(\bar\omega) z^2}{\bar\omega(z^2+\bar\omega^2)}d\bar\omega\,,
\end{align}
which can for some spectral functions be explicitly evaluated.
For an initial product state $\rho_0 = \rho_S^0 \otimes \rho_B^0$ with a thermal 
reservoir $\rho_B^0 = e^{-H_B/(k_B T)}/Z$ we could directly take expectation values and solve the equations of motion 
for position $\expval{x}_t$ and momentum $\expval{p}_t$ exactly.
We are however interested also in the asymptotic long-term limit of two-point operators and therefore keep the operator structure for now. 

\subsection{Long-term limit: Existence of a bound state}\label{APP:exactsolution:existence}

The asymptotic long-term dynamics heavily depends on the analytic properties of the function $z^2+\Omega f(z)$ that occurs in the denominators of~\eqref{EQ:solution_laplace}.
Physical intuition suggests that the equation $z^2+\Omega f(z)=0$ can only have solutions with $\Re z \le 0$ as otherwise, the system would become unstable.
However, for the long-term dynamics, it is particularly relevant whether a purely imaginary solution -- the BS -- may exist.
By using e.g. the Sochotskij-Plemelj theorem
$\lim_{\epsilon\to0} \frac{1}{x\pm\ii\epsilon}={\cal P} \frac{1}{x} \mp \ii\pi\delta(x)$ we can establish the relation 
\begin{align}
\lim_{\epsilon\to 0^+} f(\epsilon+\ii\omega) &=  \lim_{\epsilon\to 0} \frac{2}{\pi} \int_0^\infty \frac{\Gamma(\bar\omega)}{\bar\omega} \frac{(\epsilon+\ii\omega)^2}{\bar\omega^2+(\epsilon+\ii\omega)^2} d\bar\omega\nn
&= \frac{2}{\pi} {\cal P} \int_0^\infty \frac{\Gamma(\bar\omega)}{\bar\omega} \frac{\omega^2}{\omega^2-\bar\omega^2}d\bar\omega + \ii \Gamma(\omega)\,.
\end{align}
Thus, when we reconsider the defining equation for the BS existence at $z=\lim_{\epsilon\to 0^+} (\epsilon+\ii\omega)$, it becomes
\begin{align}
0 = \Omega^2-\omega^2 + \frac{2\Omega}{\pi} {\cal P} \int_0^\infty \frac{\Gamma(\bar\omega)}{\bar\omega} \frac{\omega^2}{\omega^2-\bar\omega^2}d\bar\omega
+ \ii \Omega \Gamma(\omega)\,.
\end{align}
As both real and imaginary parts have to vanish separately, spectral functions without band gaps exclude the existence of BSs.
In contrast, for our example~\eqref{EQ:sfrubin}, a BS may in principle exist.

Considering now in particular the Rubin spectral function~\eqref{EQ:sfrubin}, we get for the integral
\begin{align}
    \frac{2\Omega}{\pi} \int_0^\infty d\omega \frac{\Gamma(\omega)}{\omega} \frac{z^2}{z^2+\omega^2} = \frac{\Gamma \Omega}{\omega_c^2} z^2 \left(\sqrt{1+\frac{\omega_c^2}{z^2}}-1\right)\,,
\end{align}
which holds with the exception of a branch cut along the interval $z\in\ii[-\omega_c,+\omega_c]$.
When we consider $z \to \pm\ii\omega$ we have to approach this from the side with positive real part, which yields for the real part 
\begin{align}
    \frac{2\Omega}{\pi} {\cal P}\int_0^\infty \frac{\Gamma(\bar\omega)}{\bar\omega} \frac{\omega^2}{\omega^2-\bar\omega^2} = 
    \frac{\Gamma \Omega}{\omega_c^2} \omega \left(\omega-\sqrt{\omega^2-\omega_c^2}\right)
    \,.
\end{align}
The conditions for the BS existence at $\omega_b$ are that both imaginary and real parts vanish separately
\begin{align}
    \omega_b&>\omega_c\,,\nn
    0 &= \Omega^2-\omega_b^2 + \Gamma\Omega \frac{\omega_b^2}{\omega_c^2}\left(1-\sqrt{1-\frac{\omega_c^2}{\omega_b^2}}\right)\,,
\end{align}
where we have now omitted the principal value as the pole at $\omega_b$ is outside the integration region for $\omega_b>\omega_c$.
This now allows for an analytic real solution for $\omega_b>\omega_c$ for specific parameters obeying~\eqref{EQ:boundstatecondition}
that is given in~\eqref{EQ:prediction_boundstate}.

\subsection{Long-term limit: Single-point operators}\label{APP:exactsolution:singlepoint}

In this section, we will always assume that the BS exists, i.e., that there is a single real solution $\omega_b$ for which 
\begin{align}
\Omega f(\pm\ii\omega_b) = \omega_b^2
\end{align}
holds. 
The simplified solution where this is not the case can by retrieved by simply leaving out the BS terms.
Once a functional form for $f(z)$ is established in~\eqref{EQ:solution_laplace}, we may obtain the full time-dependent solution.
For simplicity, we only discuss the asymptotic long-term limit here and therefore only keep poles $z_i$ with a vanishing real part.
This allows us to compute the inverse Laplace transforms as
\begin{align}
g(t) &= \lim_{t\to\infty} {\cal L}^{-1}\frac{z}{z^2+\Omega f(z)}\nn
&= \mathop{\mathrm{Res}}\limits_{z=+\ii\omega_b} \frac{z e^{+z t}}{z^2+\Omega f(z)} + \mathop{\mathrm{Res}}\limits_{z=-\ii\omega_b} \frac{z e^{+z t}}{z^2+\Omega f(z)}\nn
&=\lim_{z\to\ii\omega_b} \frac{z(z-\ii\omega_b) e^{+z t}}{z^2+\Omega f(z)} + \lim_{z\to-\ii\omega_b} \frac{z(z+\ii\omega_b) e^{+z t}}{z^2+\Omega f(z)}\nn
&= \frac{1}{2} \frac{e^{+\ii\omega_b t}}{1 + \frac{\Omega f'(+\ii\omega_b)}{2(+\ii)\omega_b}}+\frac{1}{2} \frac{e^{+\ii\omega_b t}}{1 + \frac{\Omega f'(-\ii\omega_b)}{2(-\ii)\omega_b}}\nn
&= \frac{\cos(\omega_b t)}{1+\frac{\Omega\bar f}{2\omega_b}}\,,
\end{align}
where we have used the Residue formula, the rule of l'H\^{o}pital and in the last step that $\bar f = \frac{f'(\pm\ii\omega_b)}{\pm\ii}$. 
Analogously, we obtain
\begin{align}
\lim_{t\to\infty} {\cal L}^{-1} \frac{\Omega}{z^2+\Omega f(z)}&= \frac{\Omega}{\omega_b} \frac{\sin(\omega_b t)}{1+\frac{\Omega\bar f}{2\omega_b}} = \frac{\Omega}{\omega_b} h(t)\,,\nn
\lim_{t\to\infty} {\cal L}^{-1} \frac{f(z)}{z^2+\Omega f(z)}&= \frac{\omega_b}{\Omega} \frac{\sin(\omega_b t)}{1+\frac{\Omega\bar f}{2\omega_b}}= \frac{\omega_b}{\Omega} h(t)\,,\nn
\lim_{t\to\infty} {\cal L}^{-1} \frac{\Omega}{[z^2+\Omega f(z)](z+\ii\omega_k)} &= \frac{\Omega[-g(t)+\ii\frac{\omega_k}{\omega_b} h(t)]}{\omega_b^2-\omega_k^2}\nn
&\qquad-\frac{\Omega e^{-\ii\omega_k t}}{\omega_k^2-\Omega f(-\ii\omega_k)}\,,\nn
\lim_{t\to\infty} {\cal L}^{-1} \frac{z}{[z^2+\Omega f(z)](z+\ii\omega_k)} &= \frac{\ii \omega_k g(t) + \omega_b h(t)}{\omega_b^2-\omega_k^2}\nn
&\qquad+\frac{\ii\omega_k e^{-\ii\omega_k t}}{\omega_k^2-\Omega f(-\ii\omega_k)}\,,
\end{align}
and the corresponding expressions for $\omega_k\to-\omega_k$.
Now, inserting this into~\eqref{EQ:solution_laplace}, we obtain the asymptotic long-term evolution of position and momentum operators
\begin{align}
    \f{x}_\infty &= g(t) x + \frac{\Omega}{\omega_b} h(t) p\\
    &\qquad+ \sum_{k} 2h_{k} b_{k}\times\nn
    &\qquad\times\left[\frac{\Omega e^{-\ii\omega_{k}t}}{\omega_{k}^2-\Omega f(-\ii\omega_{k})}
    +\frac{\Omega g(t)-\ii\frac{\Omega \omega_k}{\omega_b} h(t)}{\omega_b^2-\omega_{k}^2} \right]\nn
    &\qquad+\sum_{k} 2h_{k}^* b_{k}^\dagger\times\nn
    &\qquad\times\left[\frac{\Omega e^{+\ii\omega_{k}t}}{\omega_{k}^2-\Omega f(+\ii\omega_{k})}
    +\frac{\Omega g(t) +\ii\frac{\Omega \omega_k}{\omega_b} h(t)}{\omega_b^2-\omega_{k}^2}\right]\,,\nn
    \f{p}_\infty &= g(t) p - \frac{\omega_b}{\Omega} h(t) x\nn
    &\qquad+ \sum_{k} 2h_{k} b_{k} \times\nn
    &\qquad\times\left[\frac{-\ii\omega_{k}e^{-\ii\omega_{k}t}}{\omega_{k}^2-\Omega f(-\ii\omega_{k})}
    -\frac{\omega_b h(t)+\ii \omega_{k} g(t) }{\omega_b^2-\omega_{k}^2}\right]\nn
    &\qquad+\sum_{k}2h_{k}^* b_{k}^\dagger\times\nn
    &\qquad\times\left[\frac{+\ii\omega_{k}e^{+\ii\omega_{k}t}}{\omega_{k}^2-\Omega f(+\ii\omega_{k})}
    -\frac{\omega_b h(t) - \ii \omega_{k} g(t)}{\omega_b^2-\omega_{k}^2}\right]\,.\nonumber
\end{align}
Taking expectation values with the assumption of an initially thermal reservoir $\expval{b_{k}}_{\rm th} = \expval{b_{k}^\dagger}_{\rm th}=0$
then shows that the first equation corresponds to the first line of Eq.~\eqref{EQ:xsolution_time}.
As a sanity check we found it convenient to verify the conservation of the commutator in the long-term limit
\begin{align}
[\f{x}_\infty, \f{p}_\infty] &= 2\ii [g^2(t)+h^2(t)]\left[1+\int\frac{d\omega}{2\pi} \frac{4\Omega\omega\Gamma(\omega)}{(\omega_b^2-\omega^2)^2}\right]\nn
&\qquad+2\ii \int_0^{\omega_c} \frac{d\omega}{2\pi} \frac{4\omega\Omega\Gamma(\omega)}{\abs{\omega^2-\Omega f(-\ii\omega)}^2}\,,
\end{align}
which is obviously not only time-independent but always assumes the value $2\ii$ (in parameter regimes where the BS does not exist, simply use $h^2(t)+g^2(t)\to 0$).

Analogously, we could write down the solution for the reservoir operators (omitted for brevity).

\subsection{Long-term limit: Two-point operators}\label{APP:exactsolution:twopoint}

To calculate the asymptotic long-term limit of two-point operators, we compute products of the expressions in the previous subsection and then take expectation values with respect to an initial product state with the reservoir taken in thermal equilibrium.
For the second moment of the position operator we obtain
\begin{align}
\expval{\f{x}^2}_\infty &= \expval{\left[g(t) x+\frac{\Omega}{\omega_b} h(t) p\right]^2}_0\nn
&\qquad +\sum_k 4 \abs{h_k}^2 [1+2 n(\omega_k)]\times\nn
&\qquad\times \abs{\frac{\Omega e^{-\ii\omega_{k}t}}{\omega_{k}^2-\Omega f(-\ii\omega_{k})}
    +\frac{\Omega g(t)-\ii\frac{\Omega \omega_k}{\omega_b} h(t)}{\omega_b^2-\omega_{k}^2}}^2\nn
&\to \expval{\left[g(t) x+\frac{\Omega}{\omega_b} h(t) p\right]^2}\nn
    &\qquad+\int_0^\infty \frac{d\omega}{2\pi} 4 \Gamma(\omega) [1+2 n(\omega)]\times\nn
    &\qquad\times
    \Omega^2\Big[\frac{1}{[\omega^2-\Omega f(-\ii\omega)][\omega^2-\Omega f(+\ii\omega)]}\nn
    &\qquad\qquad +\frac{g^2(t) +\frac{\omega^2}{\omega_b^2} h^2(t)}{(\omega_b^2-\omega^2)^2}\Big]\,,
\end{align}
where we have used the Riemann-Lebesgue lemma (Fourier transforms of $L^1$-functions vanish at infinity) that leads to the vanishing of the cross term in the continuum and long-term limit.
This reproduces the second line of~\eqref{EQ:xsolution_time}.

In an analogous fashion, we obtain the momentum variance in the asymptotic long-term and continuum limits
\begin{align}
\expval{\f{p}^2}_\infty &\to \expval{\left[g(t) p -\frac{\omega_b}{\Omega} h(t) x\right]^2}_0\nn
&\qquad+\int_0^\infty \frac{d\omega}{2\pi} 4 \Gamma(\omega) [1+2 n(\omega)]\times\nn
    &\qquad\times
    \omega^2\Big[\frac{1}{[\omega^2-\Omega f(-\ii\omega)][\omega^2-\Omega f(+\ii\omega)]}\nn
    &\qquad\qquad + \frac{g^2(t) +\frac{\omega_b^2}{\omega^2} h^2(t)}{(\omega_b^2-\omega^2)^2}\Big]\,.
\end{align}
Together, these also yield the occupation via $a^\dagger a  = (x^2+p^2)/4-1/2$.

\subsection{Long-term and weak-coupling limit}\label{APP:exactsolution:weak}

In this limit, we assume that $\Omega\in(0,\omega_c)$ and small $\Gamma$ such that~\eqref{EQ:boundstatecondition} is not obeyed.
We make use of the well-known Dirac-$\delta$-function representation
\begin{align}\label{EQ:diracdeltarep}
\pi \delta(x) = \lim_{\epsilon\to 0} \frac{\epsilon}{x^2+\epsilon^2}\,.
\end{align}
Then, we may rewrite the factors in the integrands of the second moments in position and momentum operators
\begin{align}
F(\omega) &= \lim_{\Gamma\to 0}\frac{4\Omega^2\Gamma(\omega)}{[\omega^2-\Omega f(-\ii\omega)][\omega^2-\Omega f(+\ii\omega)]}\\
&= \lim_{\Gamma\to 0} \frac{4\Omega^2 \Gamma(\omega)}{\left[\omega^2-\Omega^2-\frac{2\Omega}{\pi} {\cal P} \int \frac{\Gamma(\bar\omega)}{\bar\omega} \frac{\omega^2}{\omega^2-\bar\omega^2}\right]^2+\Omega^2\Gamma^2(\omega)}\nn
&= 4\pi\Omega\delta(\omega^2-\Omega^2) = 2\pi[\delta(\omega-\Omega)-\delta(\omega+\Omega)]\,.\nonumber
\end{align}
Plugging this in the long-term limits of the second moments, we find that the system thermalizes with the reservoir temperature
$\expval{\f{x}^2}_\infty \to 1+2n(\Omega)$ and $\expval{\f{p}^2}_\infty \to 1+2n(\Omega)$.
Note that this also implies that $\expval{a^\dagger a}_\infty \to n(\Omega)$.

\section{RC mapping}\label{APP:rcmapping}

\subsection{Single RC mapping}\label{APP:rcmapping:single}

We derive the mapping for a generic coupling operator (in the main text we use $S=a+a^\dagger$), demonstrating that the nature of the system is arbitary.
We want to equate two representations of the same Hamiltonian (we absorb phases of the amplitudes in the reservoir operators and assume $h_k,H_k \in \mathbb{R}$)
\begin{align}
H &= H_S + \sum_k \omega_k \left(b_k^\dagger + \frac{h_k}{\omega_k} S\right)\left(b_k + \frac{h_k}{\omega_k} S\right)\\
&= H_S + \Omega_1 \left(B_1^\dagger+\frac{\lambda_1}{\Omega_1} S\right)\left(B_1+\frac{\lambda_1}{\Omega_1} S\right)\nn
&\qquad+ \sum_{q>1} \Omega_q B_q^\dagger B_q + (B_1+B_1^\dagger) \sum_q H_q (B_q+B_q^\dagger)\,.\nonumber
\end{align}
In the last line we could also add a counter term of the form $\sum_q \frac{H_q^2}{\Omega_q} (B_1+B_1^\dagger)^2$ to make it manifestly positive definite.
This would eventually not change the net mapping relation, such that we omit it for brevity.
We demand that the interaction term $H_I= S \sum_k h_k (b_k + b_k^\dagger) = \lambda_1 S (B_1+B_1^\dagger)$ and also the counter term for the energy 
$\Delta H_S = \sum_k \frac{h_k^2}{\omega_k} S^2 = \frac{\lambda_1^2}{\Omega_1} S^2$ are the same.
This yields the relations
\begin{align}\label{EQ:rcrelations}
\frac{\lambda_1^2}{\Omega_1} &= \sum_k \frac{h_k^2}{\omega_k}\,,\nn
\lambda_1 (B_1+B_1^\dagger) &= \sum_k h_k (b_k + b_k^\dagger)\,,
\end{align}
which can be fulfilled for the Bogoliubov transform
\begin{align}
b_k &= \sum_q \Lambda_{kq} \frac{1}{2} \left(\sqrt{\frac{\omega_k}{\Omega_q}} + \sqrt{\frac{\Omega_q}{\omega_k}}\right) B_q\nn
&\qquad+\sum_q \Lambda_{kq} \frac{1}{2} \left(\sqrt{\frac{\omega_k}{\Omega_q}} - \sqrt{\frac{\Omega_q}{\omega_k}}\right) B_q^\dagger\,,
\end{align}
when we fix the first row of the otherwise unspecified orthogonal transform $\Lambda_{kq}$ as $\Lambda_{k1} = \frac{h_k}{\lambda_1} \sqrt{\frac{\omega_k}{\Omega_1}}$.
Inserting this in~\eqref{EQ:rcrelations} then yields in the continuum limit the RC energy and coupling strength
\begin{align}\label{EQ:rcmapping1a}
\Omega_1^2 &= \frac{\int_0^\infty d\omega \omega \Gamma(\omega)}{\int_0^\infty d\omega \frac{\Gamma(\omega)}{\omega}}\,,\nn
\lambda_1^2 &= \frac{1}{2\pi\Omega_1}\int_0^\infty d\omega \omega \Gamma(\omega)\,,
\end{align}
and the first two Eqns. of~\eqref{EQ:rcmapping2} follow directly when $\Gamma(\omega)$ only has support on a small interval.

The derivation of the spectral function mapping is analogous to Ref.~\cite{nazir2019a}, only generalized by the inclusion of the counter-term, and utilizes the Heisenberg equations of motion 
to find a relation between the spectral functions.
In the original representation, they read for a generic hermitian system observable $A=A^\dagger$ 
\begin{align}
\f{\dot{A}} &= \ii \f{S_1} + \ii \f{S_2} \sum_k h_k (\f{b_k} + \f{b_k^\dagger})\,,\nn
\f{S_1} &= [\f{H_S}+\f{\Delta H_S}, \f{A}]\,,\qquad \f{S_2} = [\f{S}, \f{A}]\,,\nn
\f{\dot{b}_k} &= -\ii \omega_k \f{b_k} -\ii h_k \f{S}\,,\qquad 
\f{\dot{b}_k^\dagger} = +\ii \omega_k \f{b_k^\dagger} + \ii h_k \f{S}\,.
\end{align}
We now Fourier-transform these equations according to $\int [\ldots] e^{+\ii z t} dt$ with the convention $\Im z > 0$.
In $z$-space, the creation and annihilation operators are no longer adjoint to each other, but we will keep the $\dagger$-notation.
This eliminates the derivatives but via the convolution theorem introduces an integral on the r.h.s. 
\begin{align}
\ii z A(z) &= \ii S_1(z) + \frac{\ii}{2\pi} \int S_2(z')\times\\
&\qquad\times \sum_k \left[h_k b_k(z-z') + h_k b_k^\dagger(z-z')\right] dz'\,,\nn
\ii z b_k(z) &= -\ii \omega_k b_k(z) -\ii h_k S(z)\,,\nn
\ii z b_k^\dagger(z) &= +\ii \omega_k b_k^\dagger(z) + \ii h_k S(z)\,.\nonumber
\end{align}
We can solve the last two equations for $b_k(z)$ and $b_k^\dagger(z)$ and
insert them into the first
\begin{align}\label{EQ:rccomp}
z A(z) &= S_1(z)+ \frac{1}{2\pi} \int dz'S_2(z') \times\\
&\qquad\times\sum_k \left[\frac{-\abs{h_k}^2}{z-z'+\omega_k} + \frac{+\abs{h_k}^2}{z-z'-\omega_k}\right] S(z-z')\nn
&= S_1(z) + \frac{1}{2\pi} \int dz' S_2(z')\times\nn
&\qquad\times\left[\frac{1}{\pi} \int_0^\infty \frac{\omega \Gamma(\omega)}{(z-z')^2-\omega^2} d\omega\right] S(z-z') \nn
&= S_1(z) - \frac{1}{2\pi} \int S_2(z') \frac{1}{2} W(z-z') S(z-z') dz'\,.\nonumber
\end{align}
Above, we have introduced the Cauchy transform of the spectral function
\begin{align}\label{EQ:cauchytrans}
W(z) = \frac{2}{\pi} \int_0^\infty \frac{\omega \Gamma(\omega)}{\omega^2-z^2} d\omega = \frac{1}{\pi} \int_{-\infty}^{+\infty} \frac{\Gamma(\omega)}{\omega-z} d\omega\,,
\end{align}
where the last equality can be obtained by splitting into partial fractions and holds for analytic continuation as an odd function $\Gamma(-\omega)=-\Gamma(+\omega)$.
In particular, we note that the spectral function can with~\eqref{EQ:diracdeltarep} be recovered from its Cauchy transform 
\begin{align}
\Gamma(\omega) = \lim_{\epsilon\to 0^+} \Im W(\omega+\ii\epsilon)\,.
\end{align}
Similarly, we can derive the Heisenberg equations of motion in the mapped representation, and Fourier-transform them according to the
same prescription, yielding
\begin{align}
z A(z) &= S_1(z) + \frac{\lambda_1}{2\pi} \int S_2(z')\times\nn
&\qquad\times \left[B_1(z-z') + B_1^\dagger(z-z')\right] dz'\,,\nn
z B_1(z) &= -\lambda_1 S(z) - \Omega_1 B_1(z)\nn
&\qquad- \sum_k H_k \left[B_k(z) + B_k^\dagger(z)\right]\,,\nn
z B_1^\dagger(z) &= +\lambda_1 S(z) + \Omega_1 B_1^\dagger(z)\nn
&\qquad + \sum_k H_k \left[B_k(z) +B_k^\dagger(z)\right]\,,\nn
z B_k(z) &= -\Omega_k B_k(z) - H_k \left[B_1(z) + B_1^\dagger(z)\right]\,,\nn
z B_k^\dagger(z) &= +\Omega_k B_k^\dagger(z) + H_k \left[B_1(z) + B_1^\dagger(z)\right]\,.
\end{align}
A potential counter term in the RC representation would lead to additional terms in the equations for $B_1(z)$ and $B_1^\dagger(z)$.
Again, we follow the approach of successively eliminating the $B_k(z)$, $B_k^\dagger(z)$, and then the 
$B_1(z)$, $B_1^\dagger(z)$ variables, eventually yielding
\begin{align}
B_1(z)+B_1^\dagger(z) = \frac{2\Omega_1\lambda_1 S(z)}{z^2-\Omega_1^2+\Omega_1 W_1(z)}\,,
\end{align}
where $W_1(z)$ denotes the Cauchy transform of the residual spectral function $\Gamma_1(\omega)$ in full analogy to~\eqref{EQ:cauchytrans}.
Inserting this in the remaining equation we obtain for the system observable
\begin{align}
z A(z) &= S_1(z) + \frac{1}{2\pi} \int S_2(z')\times\\
&\quad\times \frac{2 \lambda_1^2 \Omega_1}{(z-z')^2 - \Omega_1^2 + \Omega_1 W_1(z-z')} S(z-z') dz'\,.\nonumber
\end{align}
As this has to match for all observables with the original representation~\eqref{EQ:rccomp}, we can infer a relation between $W(z-z')$ and $W_1(z-z')$, which 
can be solved for the latter to eventually obtain the residual spectral density
\begin{align}\label{EQ:rcmapping1b}
\Gamma_1(\omega) &= -\lim_{\epsilon\to 0^+} \Im \frac{4\lambda_1^2}{W(\omega+\ii\epsilon)}\nn
&= \frac{+4 \lambda_1^2 \Gamma(\omega)}{\left[\frac{1}{\pi}{\cal P} \int_{-\infty}^{+\infty} \frac{\Gamma(\omega')}{\omega-\omega'} d\omega'\right]^2
+ \left[\Gamma(\omega)\right]^2}\,,
\end{align}
which directly yields the last of~\eqref{EQ:rcmapping2} when the support of $\Gamma(\omega)$ is restricted to finite intervals.

\subsection{Rubin spectral function and single RC}\label{APP:rcmapping:rubin}

For the Rubin example~\eqref{EQ:sfrubin} and a single RC we would obtain from the above mapping 
$\Omega_1=\frac{\omega_c}{2}$, $\lambda_1 = \frac{\sqrt{\Gamma\omega_c}}{4}$ and
\begin{align}
\Gamma_1(\omega) = \omega \sqrt{1-\frac{\omega^2}{\omega_c^2}} \Theta(\omega)\Theta(\omega_c-\omega)\,,
\end{align}
i.e., up to a constant the Rubin spectral function does not change under the mapping.
This transformed spectral function is upper-bounded $\Gamma_1(\omega) \le \omega_c/2$, which would allow for a perturbative treatment 
of strong-coupling scenarios within the supersystem (large $\Gamma$) provided that one has a small bandwidth reservoir (small $\omega_c$).
Further recursive mappings would transform the reservoir into a chain, but the above transformed spectral function already corresponds to the limiting case under the mapping~\eqref{EQ:rcmapping1b}, so a perturbative treatment would still fail for large $\omega_c$.
Nevertheless, one may use truncated chain representations for finite-time simulations.

The supersystem excitation matrix~\eqref{EQ:excitationmatrix} would become
\begin{align}
M = \left(\begin{array}{cc}
\Omega^2 + \frac{\Gamma\Omega}{2} & \sqrt{\frac{\Gamma\Omega\omega_c^2}{8}}\\
\sqrt{\frac{\Gamma\Omega\omega_c^2}{8}} & \frac{\omega_c^2}{4}
\end{array}\right)\,,
\end{align} 
and its eigenvalues can be computed analytically and already show the anticipated behaviour (the triangle symbols in Fig.~\ref{FIG:boundstatewithrc} show the larger one $\epsilon_2^2$).

When we further assume the original system energy well inside the band $\Omega = \frac{\omega_c}{2}$, we obtain for the eigenvalues and the orthogonal matrix
diagonalizing $M$
\begin{align}\label{EQ:orthogonaltrafo}
\Lambda^{\rm T} M \Lambda &= \left(\begin{array}{cc}
\epsilon_1^2\\
& \epsilon_{2}^2
\end{array}\right)\,,
\end{align}
the expressions
\begin{align}
\epsilon_1^2 &= \frac{\omega_c^2}{4}\left(1+\frac{1}{2}\frac{\Gamma}{\omega_c} - \frac{1}{2}\frac{\Gamma}{\omega_c}\sqrt{1+4\frac{\omega_c}{\Gamma}}\right)\,,\nn
\epsilon_2^2 &= \frac{\omega_c^2}{4}\left(1+\frac{1}{2}\frac{\Gamma}{\omega_c} + \frac{1}{2}\frac{\Gamma}{\omega_c}\sqrt{1+4\frac{\omega_c}{\Gamma}}\right)\,,\nn
\Lambda_{11} &= \frac{1-\sqrt{1+4\frac{\omega_c}{\Gamma}}}{\sqrt{2}\sqrt{1+4\frac{\omega_c}{\Gamma}-\sqrt{1+4\frac{\omega_c}{\Gamma}}}}\,,\nn
\Lambda_{12} &= \frac{-1-\sqrt{1+4\frac{\omega_c}{\Gamma}}}{\sqrt{2}\sqrt{1+4\frac{\omega_c}{\Gamma}+\sqrt{1+4\frac{\omega_c}{\Gamma}}}}\,.
\end{align}
In the weak-coupling regime $\Gamma \ll \omega_c$, this leads to fast mode mixing as $\Lambda_{11},\Lambda_{12} \to -1/\sqrt{2}$.
In contrast, for strong couplings we obtain $\Lambda_{11}\to 0$ and $\Lambda_{12}\to -1$, such that mode-mixing is reduced.
Altogether, we may expand for strong couplings the supersystem operators in the supersystem eigenmodes as
\begin{align}\label{EQ:bogoliubov1}
a+a^\dagger &= \sum_q \Lambda_{1q} \sqrt{\frac{\Omega}{\epsilon_q}} (c_q + c_q^\dagger)\nn
&\approx- \frac{\omega_c^{1/4}}{\Gamma^{1/4}}(c_1+c_1^\dagger+c_2+c_2^\dagger)+\ord\left\{\frac{\omega_c^{5/4}}{\Gamma^{5/4}}\right\}\,,\nn
B_1+B_1^\dagger &= \sum_q \Lambda_{2q} \sqrt{\frac{\Omega_1}{\epsilon_q}} (c_q+c_q^\dagger)\nn
&\approx \frac{\Gamma^{1/4}}{\omega_c^{1/4}} (c_1+c_1^\dagger)+\ord\left\{\frac{\omega_c^{3/4}}{\Gamma^{3/4}}\right\}\,.
\end{align}

\subsection{Weak variation expansion}\label{APP:rcmapping:weakvariation}

We may deliberately partition the reservoir modes into sub-reservoirs according to their energy, and introduce a spectral function for each sub-reservoir, that then has support 
over the energy range of that sub-reservoir only.
For such a spectral function non-vanishing in the interval ${\cal I}_i=[\omega_a,\omega_b]$, the principal value integral in the RC mapping~\eqref{EQ:rcmapping1b} becomes
\begin{align}
{\cal P} \int \frac{\Gamma(\omega')}{\omega'-\omega} d\omega' &= {\cal P} \int\limits_{\omega_a}^{\omega_b} d\omega' \frac{\Gamma(\omega')}{\omega'-\omega} +
\int\limits_{-\omega_b}^{-\omega_a} \frac{\Gamma(\omega')}{\omega'-\omega} d\omega'\nn
&= \Gamma(\omega) \ln\left(\frac{\omega_b-\omega}{\omega-\omega_a}\right)\nn
&\qquad+ \int_{\omega_a}^{\omega_b} d\omega' \left[\frac{\Gamma(\omega')-\Gamma(\omega)}{\omega'-\omega}+ \frac{\Gamma(\omega')}{\omega'+\omega}\right]\nn
&\approx \Gamma(\omega) \ln\left(\frac{\omega_b-\omega}{\omega-\omega_a}\right)
+ \Gamma(\omega) \frac{\omega_b-\omega_a}{\omega+\omega_a}\nn
&\qquad + \Gamma'(\omega)(\omega_b-\omega_a)+\ldots
\end{align}
In the first equality, we have made the support in the intervals and the continuation to the real axis explicit,
in the second we have inserted $\Gamma(\omega')=\Gamma(\omega)+\Gamma(\omega')-\Gamma(\omega)$ to separate the divergent parts, and 
in the last approximation we have expanded $\Gamma(\omega')$ around $\Gamma(\omega)$.
Now, assuming a small width $\omega_b-\omega_a$ we would get from~\eqref{EQ:rcmapping1a} that 
$\lambda_1^2 \approx \frac{\Gamma(\Omega_1)(\omega_b-\omega_a)}{2\pi}$ and consequently
\begin{align}
\Gamma_1(\omega) \approx \frac{2}{\pi} \frac{\omega_b-\omega_a}{1+\frac{1}{\pi^2} \ln^2\left(\frac{\omega_b-\omega}{\omega-\omega_a}\right)}
\le \frac{2(\omega_b-\omega_a)}{\pi}\,,
\end{align}
which becomes small when the interval width $\omega_b-\omega_a$ is small.
By introducing many RCs we may however reach arbitrarily fine discretization such that we will eventually be allowed to neglect the higher-order terms.
This allows to upper-bound the residual spectral function for sufficiently many RCs by $\Gamma_i(\omega) \lesssim \frac{2}{\pi}\Delta\omega_i$,
which corresponds to the dashed line in Fig.~\ref{FIG:rcmappingspecdens} right panel.

\subsection{Supersystem excitation energies}\label{APP:rcmapping:excitations}

Even in case of a harmonic system $\Omega=a^\dagger a$ one is left with the task of finding the supersystem excitation energies, i.e., the eigenvalues of $\tilde H_S$ that is given by the 
first line of Eq.~\eqref{EQ:model3}.
We can represent it with the rescaled position $\{\tilde X, \tilde X_i\}$ and momentum operators $\{\tilde P,\tilde P_i\}$ via $a=\sqrt{\frac{\Omega}{2}} \tilde X + \frac{\ii}{\sqrt{2\Omega}} \tilde P$ and $B_i=\sqrt{\frac{\Omega_i}{2}} \tilde X_i + \frac{\ii}{\sqrt{2\Omega_i}} \tilde P_i$ as
\begin{align}
\tilde H_S &=\frac{1}{2} \tilde P^2+\frac{1}{2}\sum_i \tilde P_i^2 + \tilde V  -\frac{\Omega}{2}-\sum_i \frac{\Omega_i}{2}\,,\nn
\tilde V  &=\frac{1}{2}\left(\Omega^2+\sum_i \frac{4\Omega\lambda_i^2}{\Omega_i}\right) \tilde X^2+\frac{1}{2}\sum_i\Omega_i^2 \tilde X_i^2\nn
&\qquad+\frac{1}{2} \sum_i 4 \lambda_i\sqrt{\Omega\Omega_i} \tilde X \tilde X_i\,.
\end{align}
The kinetic part is invariant with respect to rotations, such that the (squared) excitation energies can be found by diagonalizing the potential part with a suitable rotation transform.
Representing the potential term $\tilde V$ as a quadratic form then directly maps to the eigenvalue problem of the matrix~\eqref{EQ:excitationmatrix}.
Formally, the representation of the supersystem Hamiltonian as $\tilde H_S = \sum_q \epsilon_q c_q^\dagger c_q + {\rm const}$  corresponds to a Bogoliubov transform.
When we denote the matrix elements of the orthogonal transformation diagonalizing~\eqref{EQ:excitationmatrix} by $\Lambda$ as in~\eqref{EQ:orthogonaltrafo},
we can represent
\begin{align}
a &= \sum_q \Lambda_{1q} \Big[\frac{1}{2} \Big(\sqrt{\frac{\Omega}{\epsilon_q}}+\sqrt{\frac{\epsilon_q}{\Omega}}\Big) c_q\nn
&\qquad+ \frac{1}{2} \Big(\sqrt{\frac{\Omega}{\epsilon_q}}-\sqrt{\frac{\epsilon_q}{\Omega}}\Big) c_q^\dagger\Big]\,,
\end{align}
which explicitly demonstrates that this transform does not preserve the quasiparticle number.

Using our specific example~\eqref{EQ:sfrubin} with the partitioning~\eqref{EQ:discretization} in the regime $\Gamma\Omega/N \gg \omega_c$, we can approximate the excitation matrix~\eqref{EQ:excitationmatrix} as
\begin{align}
M \approx \sqrt{\frac{\Gamma\Omega\omega_c^2}{8 N}} \left(\begin{array}{cccc}
\sqrt{\frac{2N \Gamma\Omega}{\omega_c^2}} & 1 & \hdots & 1\\
1 & 0 & \hdots & 0\\
\vdots & \vdots & & \vdots\\
1 & 0 & \hdots & 0
\end{array}\right)\,,
\end{align}
which allows to represent in the limit of strong couplings the corresponding eigenvector as
\begin{align}
\Lambda_{1,q} \approx \left(\frac{\omega_c}{\sqrt{2N\Gamma\Omega}},\ldots,\frac{\omega_c}{\sqrt{2N\Gamma\Omega}},1-\frac{\omega_c^2}{4\Gamma\Omega}\right)\,,
\end{align}
i.e., it couples the original system dominantly to the BS mode and weakly to the $N$ band modes
\begin{align}\label{EQ:bogoliubov}
a &\approx \sum_{q=1}^N \frac{\omega_c}{\sqrt{2N \Gamma\Omega}} \Big[\frac{1}{2} \Big(\sqrt{\frac{\Omega}{\epsilon_q}}+\sqrt{\frac{\epsilon_q}{\Omega}}\Big) c_q\nn
&\qquad+ \frac{1}{2} \Big(\sqrt{\frac{\Omega}{\epsilon_q}}-\sqrt{\frac{\epsilon_q}{\Omega}}\Big) c_q^\dagger\Big]\nn
&\qquad+\left(1-\frac{\omega_c^2}{4\Gamma\Omega}\right) \Big[\frac{1}{2} \Big(\sqrt{\frac{\Omega}{\epsilon_{N+1}}}+\sqrt{\frac{\epsilon_{N+1}}{\Omega}}\Big) c_{N+1}\nn
&\qquad+ \frac{1}{2} \Big(\sqrt{\frac{\Omega}{\epsilon_{N+1}}}-\sqrt{\frac{\epsilon_{N+1}}{\Omega}}\Big) c_{N+1}^\dagger\Big]\,.
\end{align}
This already shows that mode mixing can be suppressed by increasing $\Gamma$, which would be beneficial for the BS lifetime.
In App.~\ref{APP:perturbative} we show that for strong $\Gamma$ when $\epsilon_{N+1}$ is inside the band gap (BS regime), the BS mode $c_{N+1}$ is inert with respect to the dissipator.

\section{Arrowhead matrices}\label{APP:arrowhead}

\subsection{Bounds on all eigenvalues}\label{APP:arrowhead:all}

The characteristic polynomial of a generic arrowhead matrix
\begin{align}\label{EQ:arrowheadmatrix}
M = \left(\begin{array}{cccc}
\alpha & \beta_1 & \beta_2 & \ldots\\
\beta_1 & \alpha_1\\
\beta_2 & & \alpha_2\\
\vdots & & & \ddots
\end{array}\right)
\end{align}
is given by
\begin{align}\label{EQ:charpoly}
D_N(\sigma) &= (\alpha-\sigma)\prod_{i=1}^N (\alpha_i-\sigma) - \sum_{i=1}^N \prod_{j\neq i} (\alpha_j-\sigma)\beta_i^2\nn
&= \left[\prod_{i=1}^N (\alpha_i-\sigma)\right]\left[\alpha-\sigma-\sum_{i=1}^N \frac{\beta_i^2}{\alpha_i-\sigma}\right]\,.
\end{align}
Now specifically considering~\eqref{EQ:excitationmatrix} and under the assumption that $\Omega_1^2 < \Omega_2^2 < \ldots < \Omega_N^2$ (otherwise, just exchange and relabel)
we find that 
\begin{align}
    D_N(\Omega_1^2) &=-4\Omega \lambda_1^2 \Omega_1 (\Omega_2^2-\Omega_1^2) \ldots (\Omega_N^2-\Omega_1^2) < 0\,,\nn
    D_N(\Omega_2^2) &=-4\Omega \lambda_2^2 \Omega_2 (\Omega_1^2-\Omega_2^2)\times\nn
    &\qquad\times (\Omega_3^2-\Omega_2^2) \ldots (\Omega_N^2-\Omega_2^2)>0\,,\nn
    D_N(\Omega_3^2) &=-4\Omega \lambda_3^2 \Omega_3 (\Omega_1^2-\Omega_3^2)(\Omega_2^2-\Omega_3^2)\times\nn
    &\qquad\times(\Omega_4^2-\Omega_3^2)\ldots(\Omega_N^2-\Omega_3^2)<0\,,\nn
    &\vdots\nn
    D_N(\Omega_N^2) &=-4\Omega \lambda_N^2 \Omega_N (\Omega_1^2-\Omega_N^2) \ldots (\Omega_{N-1}^2-\Omega_N^2)
\end{align}
have alternating signs, which bounds the roots of $D_N(\sigma)$ between the $\Omega_i^2$-values. 
Furthermore, we have ${\rm sgn}(D_N(-\infty))=+1$ and ${\rm sgn}(D_N(+\infty))=(-1)^{N+1}=-{\rm sgn}(D_N(\Omega_N^2))$, 
which bounds the $N+1$ eigenvalues as $-\infty < \epsilon_1^2 < \Omega_1^2 < \epsilon_2^2 < \ldots < \epsilon_N^2 < \Omega_N^2 < \epsilon_{N+1}^2$, as one may also obtain via 
the Cauchy interlacing (or Poincar\'e separation) theorem.
We can furthermore bound the lowest eigenvalue by evaluating $D_N(0)$ 
\begin{align}
D_N(0) &= \alpha \prod_i \alpha_i - \sum_i \frac{\beta_i^2}{\alpha_i} \prod_j \alpha_j = \prod_j \alpha_j \left[\alpha - \sum_i \frac{\beta_i^2}{\alpha_i}\right]\nn
 &= \prod_j \Omega_j^2 \left[\Omega^2 + \sum_i \frac{4\Omega \lambda_i^2}{\Omega_i} - \sum_i \frac{4 \Omega \lambda_i^2 \Omega_i}{\Omega_i^2}\right]\nn
 &= \prod_j \Omega_j^2 \Omega^2 > 0\,,
\end{align}
which tightens the bound on the lowest eigenvalue as $0 < \epsilon_1^2$.

\subsection{Bounds on the largest eigenvalue}\label{APP:arrowhead:largest}

Furthermore, the sum of all eigenvalues equals the trace $T$ of the matrix~\eqref{EQ:excitationmatrix}
\begin{align}
T &= \Omega^2 + \sum_i \frac{4\Omega \lambda_i^2}{\Omega_i} + \sum_i \Omega_i^2 = \sum_{i=1}^N \epsilon_i^2 + \epsilon_{N+1}^2\,,
\end{align}
which allows to express the largest eigenvalue as
\begin{align}
\epsilon_{N+1}^2 = \Omega^2 + \sum_i \frac{4\Omega \lambda_i^2}{\Omega_i} + \sum_{i=1}^N [\Omega_i^2-\epsilon_i^2]\,.
\end{align}
In the last summation, all terms are positive (see above), simply neglecting them yields the loose lower bound on the eigenvalue~\eqref{EQ:boundlooselower}
that corresponds to the lower dotted blue line in Fig.~\ref{FIG:boundstatewithrc}.
The same bound could be obtained from the Cauchy interlacing theorem by cutting rows and columns from $2\ldots N$ and considering the continuum limit.

To improve this lower bound, for an arrowhead matrix~\eqref{EQ:arrowheadmatrix}, we can define the unitary matrix
\begin{align}
U = \left(\begin{array}{ccccc}
1 & 0 & \hdots & \hdots & 0\\
0 & v_{11} & \hdots & \hdots & v_{1N}\\
0 & v_{21} & \hdots & \hdots & v_{2N}\\
\vdots & \vdots &&& \vdots\\
0 & v_{N1} & \hdots & \hdots & v_{NN}
\end{array}\right)
\end{align}
with $\{\f{v_i}\}$ denoting an orthonormal set of vectors where specifically the components of $\f{v_1}$ are chosen as $v_{1j}=\beta_j/\sqrt{\sum_i \beta_i^2}\equiv \beta_j/\beta$.
This then implies that the transformed matrix
\begin{align}
U M U^\dagger =\left(\begin{array}{cc|ccc}
\alpha & \beta & 0 & \hdots & 0\\
\beta & \sum_i \frac{\beta_i^2 \alpha_i}{\beta^2} & \tilde \beta_2 & \hdots & \tilde \beta_N\\
\hline
0 & \tilde \beta_2 &\\
\vdots & \vdots & & M_{\rm red}\\
0 & \tilde \beta_N
\end{array}\right)
\end{align}
has the same eigenvalues, and from the Cauchy interlacing (or Poincar\'e separation) theorem we can conclude that the ordered eigenvalues $\sigma_1 \le \ldots \le \sigma_{N+1}$ are bounded by the ordered eigenvalues of the top left submatrix $\mu_1 \le \mu_2$ as
\begin{align}
\sigma_1 \le \mu_1 \le \sigma_N\,,\qquad
\sigma_2 \le \mu_2 \le \sigma_{N+1}\,,
\end{align}
i.e., the larger eigenvalue of the top left submatrix yields a lower bound on the largest eigenvalue of~\eqref{EQ:arrowheadmatrix}.
Specifically for our model~\eqref{EQ:excitationmatrix} we obtain
\begin{align}
\alpha &= \Omega^2 + \frac{2\Omega}{\pi} \int \frac{\Gamma(\omega)}{\omega} d\omega\,,\nn
\beta^2 &= \frac{2\Omega}{\pi} \int \omega \Gamma(\omega) d\omega\,,
\end{align}
which holds independent of any discretization resolution.
In the continuum limit, we then obtain
\begin{align}
\sum_i \frac{\beta_i^2 \alpha_i}{\beta^2} &= \frac{1}{\beta^2} \sum_i 4\Omega \lambda_i^2 \Omega_i^3
\approx \frac{4\Omega}{2\pi\beta^2} \int \omega^3 \Gamma(\omega) d\omega\nn
&= \frac{\int \omega^3 \Gamma(\omega) d\omega}{\int \omega \Gamma(\omega) d\omega} \to \frac{\omega_c^2}{2}\,,
\end{align}
where the last limit is taken with respect to our example~\eqref{EQ:sfrubin} for which we also obtain $\alpha \to \Omega^2+\frac{\Gamma\Omega}{2}$ and
$\beta\to \sqrt{\frac{\Gamma\Omega\omega_c^2}{8}}$.
The resulting bound is plotted as the violet dotted curve in Fig.~\ref{FIG:boundstatewithrc}, it is much tighter than the simple bound~\eqref{EQ:boundlooselower}.

An upper bound on the largest eigenvalue $\sigma_{\rm max} = \epsilon_{N+1}^2$ can be derived from the characteristic polynomial~\eqref{EQ:charpoly}, which for the matrix~\eqref{EQ:excitationmatrix} yields the equation
\begin{align}\label{EQ:sigmamax}
\sigma_{\rm max}-\Omega^2-\sum_i \frac{4\Omega\lambda_i^2}{\Omega_i} &= \sum_{i=1}^N \frac{4 \Omega \lambda_i^2\Omega_i}{\sigma_{\rm max}-\Omega_i^2}\nn
&\le \frac{4\Omega}{\sigma_{\rm max}-\Omega_N^2} \sum_{i=1}^N \lambda_i^2 \Omega_i\,.
\end{align}
Converting back to integral representations we obtain 
\begin{align}
\abs{\sigma_{\rm max} - \Omega^2 - \frac{2\Omega}{\pi} \int d\omega\frac{\Gamma(\omega)}{\omega}}\abs{\sigma_{\rm max}-\Omega_N^2}\nn
\le \frac{2\Omega}{\pi} \int d\omega \omega \Gamma(\omega)\,,
\end{align}
which with $\Omega_N^2\to\omega_c^2$ corresponds to the dashed red curve in Fig.~\ref{FIG:boundstatewithrc}.
The same bound can be derived for~\eqref{EQ:discretization} based on Brauer's ovals of Cassini~\cite{varga2001a}.
The above considerations also yield a lower bound by replacing $\Omega_N^2\to\Omega_1^2$ and reversing the inequality, which however is looser than the bound discussed before. 

\subsection{Critical coupling strength}\label{APP:arrowhead:critical}

We may turn the question around and ask -- at which coupling strength does a BS exist, i.e., when reaches the largest eigenvalue the band boundary?
Assuming a single band in $[0,\omega_c]$ we would find this from~\eqref{EQ:sigmamax} by replacing $\sigma_{\rm max} \to \omega_c^2$
\begin{align}
\omega_c^2 - \Omega^2 - \frac{2\Omega}{\pi} \int_0^{\omega_c} \frac{\Gamma(\omega)}{\omega} d\omega &= \frac{2\Omega}{\pi} \lim_{N\to\infty} \sum_{i=1}^N \frac{\Omega_i \Gamma(\Omega_i)}{\omega_c^2-\Omega_i^2} \Delta\Omega_i\nn
&\to \frac{2\Omega}{\pi} \int_0^{\omega_c} \frac{\omega\Gamma(\omega)}{\omega_c^2-\omega^2} d\omega\,,
\end{align}
where we have assumed that the spectral function approaches zero at least linearly at the band gap boundary.
When inserting our example~\eqref{EQ:sfrubin} we would just recover~\eqref{EQ:boundstatecondition}.
When the spectral function approaches zero sub-linearly, we may have to split off the last summation term (analogous to Bose-Einstein condensation).

\section{Perturbative treatment}\label{APP:perturbative}

In RC representation, the supersystem is only coupled via the residual coupling $\Gamma_{1\ldots N}(\omega)$ to $N$ independent sub-reservoirs of small energy range.
As to second order in the $H_{iq}$ these can be treated independently, let us consider only the coupling to one such subreservoir via the coupling operator $X_i = B_i+B_i^\dagger$ -- the total dissipator can be reconstructed later-on.
In RC representation, the total Hamiltonian can be expressed as $H=\tilde H_S + \tilde H_I + \tilde H_B$ with $\tilde H_S$ given by the first line of~\eqref{EQ:model3}.
For generality, we further split the supersystem Hamiltonian $\tilde H_S = \tilde H_S^0 + \tilde H_S^1$ into a quadratic part $\tilde H_S^0$ and a weak part anharmonic part $H_S^1$ (if interactions are present). 
In the interaction picture with respect to $\tilde H_S^0+\tilde H_B$ with system-bath coupling $\tilde H_I= A \otimes B$,
one may define the reservoir correlation function $C(\tau)=\trace{e^{+\ii \tilde H_B \tau} B e^{-\ii \tilde H_B \tau} B \tilde \rho_B}$ and write 
the Redfield-II master equation (compare e.g. Eq.~(27) of Ref.~\cite{landi2022a}) as
\begin{align}
\f{\dot \rho_S} &= -\ii [\f{\tilde H_S^1}, \f{\rho_S}]
- \int_0^\infty d\tau \Big\{C(+\tau) [\f{A}(t), \f{A}(t-\tau)\f{\rho_S}(t)]\nn
&\qquad + C(-\tau)[\f{\rho_S}(t)\f{A}(t-\tau), \f{A}(t)]\Big\}\,.
\end{align}
As $\tilde H_S^0$ does not contain interactions, we may diagonalize it $\tilde H_S^0 = \sum_q \epsilon_q c_q^\dagger c_q$ and accordingly write the system coupling operator as (we omit the index of the RC for brevity and consider only its associated bath)
\begin{align}
\f{A}(t) &= e^{+\ii \tilde H_S^0 t} (B+B^\dagger) e^{-\ii \tilde H_S^0 t}\nn
&= \sum_q \left(\alpha_q e^{-\ii\epsilon_q t} c_q + \alpha_q^* e^{+\ii\epsilon_q t} c_q^\dagger\right)\nn
&= \f{A_{\rm band}}(t) + \left[\alpha_{N+1} e^{-\ii \epsilon_{N+1}t} c_{N+1} + {\rm h.c.}\right]\,,
\end{align}
to isolate the time-dependence.
Thus, when $\epsilon_{N+1}$ is outside the band (interpreted as BS) such that $\epsilon_{N+1} \gg {\rm max}_\omega \Gamma_i(\omega)$, we may perform a partial secular approximation~\cite{farina2019a,cattaneo2019a}, which yields an LGKS generator for the BS and simply maintains the Redfield generator for the other modes.
Back in the Schr\"odinger picture, it reads
\begin{align}
\dot\rho_S 
&= -\ii [\tilde H_S^0 + \tilde H_S^1 + \Delta\epsilon_{N+1} c_{N+1}^\dagger c_{N+1}, \rho_S]\nn
&\qquad- \int_0^\infty d\tau \Big\{C(+\tau) [A_{\rm band}, e^{-\ii H_S \tau} A_{\rm band} e^{+\ii H_S \tau} \rho_S]\nn
&\qquad + C(-\tau)[\rho_Se^{-\ii H_S \tau} A_{\rm band} e^{+\ii H_S\tau}, A_{\rm band}]\Big\}\nn
&\qquad+\abs{\alpha_{N+1}}^2 \gamma(+\epsilon_{N+1}) {\cal D}_{c_{N+1}}\rho_S\nn
&\qquad+\abs{\alpha_{N+1}}^2 \gamma(-\epsilon_{N+1}) {\cal D}_{c_{N+1}^\dagger} \rho_S\,,
\end{align}
with ${\cal D}_a \rho \equiv a \rho a^\dagger - \frac{1}{2}\left\{a^\dagger a, \rho\right\}$ and where $\Delta\epsilon_{N+1}$ denotes possible Lamb-shift corrections.
The Fourier transform of the correlation function is for a bosonic reservoir with linear coupling $B=\sum_q H_q (B_q+B_q^\dagger)$ given by
\begin{align}
\gamma(\omega) = \Gamma(\omega)[1+n(\omega)]\,,
\end{align}
where as before $\Gamma(\omega)=-\Gamma(-\omega)$.
As we consider the residual reservoir here, we have $\Gamma(\omega)\to\Gamma_i(\omega)$, such that due to 
$\Gamma_i(\pm\epsilon_{N+1})=0$ the last two lines in the above equation would actually vanish.
With $[c_{N+1},A_{\rm band}]=0$ the stability of the BS under the dissipator follows immediately.
It can thus only decay when mode-mixing with the vulnerable band modes occurs -- which is only possible when $\tilde H_S^1$ contains interactions.

\end{document}